\documentclass[preprint,prd,tightenlines,floatfix,
nofootinbib,eqsecnum,superscriptaddress]{revtex4-1}

\usepackage[T1]{fontenc}		

\usepackage{amsmath,amsfonts,amssymb,amstext,mathrsfs}
\usepackage{amsthm}
\usepackage{mathpazo}

\usepackage[dvips]{graphicx}
\usepackage{epsf,float}
\usepackage{revsymb}

\usepackage{dcolumn}
\usepackage{braket}
\usepackage{color,xcolor}
\usepackage{graphicx}
\usepackage{subfigure}
\usepackage{multirow}
\usepackage{tabularx}
\usepackage{pstricks}
\usepackage[section]{placeins}
\usepackage{booktabs}
\usepackage{array}

\usepackage{tablefootnote}

\usepackage{commath}

\usepackage{hyperref}

\usepackage{lineno}
\usepackage{hyphenat}

\newcommand{\Pom}{\mathbb{P}}

\newcommand{\Reg}{\mathbb{R}}

\renewcommand\slash[1]{\not \! #1}

\usepackage[normalem]{ulem}  

\begin{document}

\nolinenumbers

\title{\boldmath 
Deeply virtual Compton scattering in the tensor-pomeron approach}

\vspace{0.6cm}

\author{Piotr Lebiedowicz}
\email{Piotr.Lebiedowicz@ifj.edu.pl}
\affiliation{Institute of Nuclear Physics Polish Academy of Sciences, 
Radzikowskiego 152, PL-31342 Krak{\'o}w, Poland}

\author{Otto Nachtmann}
\email{O.Nachtmann@thphys.uni-heidelberg.de}
\affiliation{Institut f\"ur Theoretische Physik, Universit\"at Heidelberg,
Philosophenweg 16, D-69120 Heidelberg, Germany}

\author{Antoni Szczurek
\footnote{Also at \textit{College of Natural Sciences, 
Institute of Physics, University of Rzesz{\'o}w, 
ul. Pigonia 1, PL-35310 Rzesz{\'o}w, Poland}.}}
\email{Antoni.Szczurek@ifj.edu.pl}
\affiliation{Institute of Nuclear Physics Polish Academy of Sciences, 
Radzikowskiego 152, PL-31342 Krak{\'o}w, Poland}

\begin{abstract}
The two-tensor-pomeron model proposed previously 
to describe low $x$ DIS data 
is applied to real and virtual Compton scattering on a proton.
The model includes two tensor pomerons, 
a soft and a hard one, and tensor reggeons.
We include contributions of both transverse 
and longitudinal virtual photons.
We show that this model gives 
a very good description of experimental data 
at small Bjorken $x$ on 
deeply virtual Compton scattering (DVCS) from HERA.
The reggeon exchange term is particularly
relevant for describing the real-photon-proton scattering
measured at lower $\gamma p$ energies at FNAL.
We present two fits which differ somewhat in the strength
of the hard pomeron contribution.
In both fits we find that the
interference between soft- and hard-pomeron exchange
plays an important role.
We find that in DVCS the soft-pomeron contribution is considerable 
up to $Q^{2} \sim 20$~GeV$^{2}$.
Our model allows to study the transition from 
the small-$Q^{2}$ regime,
including the photoproduction ($Q^{2} = 0$) limit,
to the large-$Q^{2}$ regime, the DIS limit.
We also discuss the ratio of cross sections
for longitudinally and transversely polarized
virtual photons in $\gamma^{*} p \to \gamma p$
as a function of $|t|$ and $Q^2$.
The ratio 
$\tilde{R}(Q^{2},W^{2},t) = (d\sigma_{\rm L} / dt) / (d\sigma_{\rm T} / dt)$
strongly increases with $t$.
Our findings may be checked
in future lepton-nucleon scattering experiments in the low-$x$ regime,
for instance, at a future Electron-Ion Collider 
at the BNL (EIC), and, if LHeC is realized, at the LHC.
\end{abstract}


\maketitle

\section{Introduction}
\label{sec:introduction}

In this Letter we apply the two-tensor-pomeron model \cite{Britzger:2019lvc}
to deeply virtual Compton scattering (DVCS) on the proton 
($\gamma^{*} p \to \gamma p$).
This Regge type model can be used
for large $\gamma^{*} p$ centre-of-mass energy 
$W \gg m_{p}, \sqrt{|t|}$, $|t| \lesssim 1$~GeV$^{2}$,
and small Bjorken-$x$, say
$x = Q^{2}/(W^{2} + Q^{2} - m_{p}^{2}) < 0.02$.
Here $m_{p}$ is the proton mass,
$Q^{2}$ is the photon virtuality, and
$t$ is the squared momentum transfer at the proton vertex.

The DVCS has been a subject of extensive experimental
and theoretical research; see e.g.
\cite{Ji:1996nm,Diehl:1997bu,Bartels:1981jh,
Diehl:2003ny, 
Diehl:2005pc,Machado:2008tp,Kopeliovich:2008ct,
Braun:2014sta,Kriesten:2019jep,Dutrieux:2021nlz}
for theory papers. For a review and many references see \cite{Diehl:2003ny}.
Recently, the possibility of probing 
the nonlinear (saturation) effects on the DVCS process 
was considered 
in \cite{Bendova:2022xhw,Xie:2022sjm,Goncalves:2022wzq}.
A comprehensive experimental overview of DVCS 
is presented in \cite{dHose:2016mda};
see also Fig.~5 of \cite{Aschenauer:2013hhw}
which shows the kinematic coverage in the $x$-$Q^{2}$ plane
for existing DVCS measurements, as well as planned ones.
The H1 \cite{H1:2005gdw,H1:2009wnw} and ZEUS
\cite{ZEUS:2003pwh,ZEUS:2008hcd} Collaborations
measured the DVCS cross section over a broad range of
$W$ and $Q^{2}$ at low Bjorken-$x$.
Also the differential cross section $d\sigma/dt$ was extracted.
In \cite{H1:2009wnw} 
the beam charge asymmetry due to the interference between 
the DVCS and the purely electromagnetic Bethe-Heitler process 
was measured.
The HERMES Collaboration \cite{HERMES:2011bou,HERMES:2012gbh,HERMES:2012idp}
performed measurements of single- and double-spin DVCS asymmetries.
The Jefferson Lab CLAS Collaboration measured
DVCS beam-spin asymmetries and cross sections
\cite{CLAS:2007clm,CLAS:2015uuo,CLAS:2018bgk,CLAS:2021gwi}
and longitudinally polarized target-spin asymmetries 
\cite{CLAS:2006krx,CLAS:2014qtk,CLAS:2015bqi}.
The COMPASS Collaboration \cite{COMPASS:2018pup}
 measured the DVCS cross section by 
studying exclusive single-photon production 
in muon-proton scattering, $\mu p \to \mu p \gamma$.
The Jefferson Lab Hall A Collaboration measured
the photon electroproduction cross section 
in the valence quark region
\cite{JeffersonLabHallA:2006prd,JeffersonLabHallA:2015dwe},
as well as the DVCS off the neutron \cite{JeffersonLabHallA:2007jdm},
and recently \cite{JeffersonLabHallA:2022pnx}
the DVCS off the proton at high values of Bjorken-$x$.
Experimental programs at Jefferson Lab,
at the Electron Ion Collider (EIC) under construction at BNL 
\cite{Accardi:2012qut,Aschenauer:2013hhw,Aschenauer:2017jsk,AbdulKhalek:2021gbh},
and, if realized, at a Large Hadron Electron Collider (LHeC) 
at the LHC \cite{LHeCStudyGroup:2012zhm,LHeC:2020van},
are expected to improve our knowledge of DVCS
in a wide kinematic range.

DVCS is a prime playground for the application
of the generalized parton-distribution (GPD) concept
based on perturbative QCD (pQCD), 
cf.~\cite{Diehl:2003ny} for a review.
However, for small values of Bjorken $x$ the application of pQCD
concepts faces the problem that the effective expansion
parameter is there $\alpha_{s} \ln(1/x)$
which is large for small $x$.
More phenomenological models are then frequently used,
based e.g. on the colour-dipole approach
or the Regge approach.
In our present work we use the latter approach where
the scattering is described using exchange objects.
For high energies $W$ the pomeron will be the most important
exchange object. For general reviews of pomeron physics
see, e.g., \cite{Collins:1977,Caneschi,Donnachie:2002en}.
Applications of Regge theory to DVCS have been given in
\cite{Capua:2006ij,Fazio:2013hza} 
and we shall comment on these papers below.


In the following we shall use the tensor-pomeron model 
as introduced for soft high-energy reactions 
in \cite{Ewerz:2013kda}
and extended to hard reactions in \cite{Britzger:2019lvc}.
The soft and hard pomeron and the charge-conjugation
$C = +1$ reggeons ($\Reg_{+} = f_{2 \Reg}, a_{2 \Reg}$)
are described as 
effective rank-2 symmetric tensor exchanges,
the odderon and the $C = -1$ reggeons
($\Reg_{-} = \rho_{\Reg}, \omega_{\Reg}$)
are described as effective vector exchanges.
Applications of this tensor-pomeron concept
were given for photoproduction of pion pairs 
in \cite{Bolz:2014mya},
for a number of exclusive central-production reactions,
see for instance 
\cite{Lebiedowicz:2013ika,Lebiedowicz:2014bea,
Lebiedowicz:2016ioh,
Lebiedowicz:2019boz,Lebiedowicz:2019jru,Lebiedowicz:2020yre},
and for soft-photon radiation (bremsstrahlung)
in hadronic collisions
\cite{Lebiedowicz:2021byo,Lebiedowicz:2022nnn}.
For some remarks on the history of tensor-pomeron concepts
and corresponding references see \cite{Ewerz:2016onn}.

Now we wish to discuss some arguments why we think 
that soft and hard pomeron should
be described as effective tensor objects.
It was shown in \cite{Britzger:2019lvc} 
that considering these pomerons as vector objects leads
to the conclusion that they decouple in the total photoabsorption
cross section on the proton and 
in the structure functions of low-$x$ deep \mbox{inelastic} scattering (DIS).
But experiment clearly shows pomeron-exchange behaviour
for these quantities at large $W$.
A scalar nature of the pomeron could be excluded
in \cite{Ewerz:2016onn}
through a comparison of spin-dependent
proton-proton elastic scattering with \mbox{theory}.
The ratio of single-flip to non-flip amplitudes
as measured in \cite{Adamczyk:2012kn} is in complete
disagreement with the prediction from the scalar pomeron
but agrees nicely with that from the tensor pomeron; 
see \cite{Ewerz:2016onn}.
A tensor nature of the soft pomeron is also preferred
in holographic-QCD approaches; 
see e.g. \cite{Domokos:2009hm,Iatrakis:2016rvj}.
A two-pomeron description of low-$x$ DIS 
was first proposed in \cite{Donnachie:1998gm}.
However, there a vector nature of the pomerons was considered
and this, at closer look, leads to the conclusion
of their decoupling in DIS; see above.

In \cite{Britzger:2019lvc} the two-tensor-pomeron approach
was applied to describe simultaneously 
the experimental data on the inclusive cross sections for DIS 
at low $x$ \cite{H1:2015ubc} and for photoproduction.
The global fit performed there was
based on the hard and soft pomeron, 
and the reggeon $f_{2 \Reg} + a_{2 \Reg}$.
The model describes the transition from the low-$Q^{2}$ regime,
including the $Q^{2} = 0$ photoproduction limit,
where the real or virtual photon behaves hadronlike
and the soft pomeron dominates,
to the large-$Q^{2}$ regime which is the domain of the hard pomeron.
For the photoproduction cross section
$\sigma_{\gamma p}(W)$
no significant contribution from the hard pomeron was found.
In this model $\sigma_{\gamma p}(W)$ is,
in the range 6~GeV~$< W < 209$~GeV,
dominated by soft-pomeron exchange with 
a significant reggeon contribution for $W < 30$~GeV.
The transition region where both pomerons
contribute significantly was found to be roughly
$0 < Q^{2} < 20$~GeV$^{2}$.
But as $Q^{2}$ increases the hard-pomeron component
becomes more and more important.

The Letter is organized as follows.
In the next section we introduce 
the tensor-pomeron approach
for real and deeply virtual Compton scattering on a proton. 
Section~\ref{sec:results} presents
the comparison of our model results
with experimental data.
There we also discuss theoretical uncertainties 
and limitations of the model.
Section~\ref{sec:conclusions} contains our conclusions.

\section{Theoretical formalism}
\label{sec:formalism}
We investigate the real and deeply virtual Compton scattering 
on a proton
\begin{equation}
\gamma^{(*)}(q, \epsilon) + p (p,\lambda) \to 
\gamma(q', \epsilon') + p (p',\lambda')\,.
\label{2.1}
\end{equation}
The momenta are indicated in brackets,
$\lambda, \lambda' \in \{1/2, -1/2 \}$
are the proton helicities,
and $\epsilon, \epsilon'$ are
the photon polarization vectors.

\begin{figure}[!h]
\includegraphics[width=7cm]{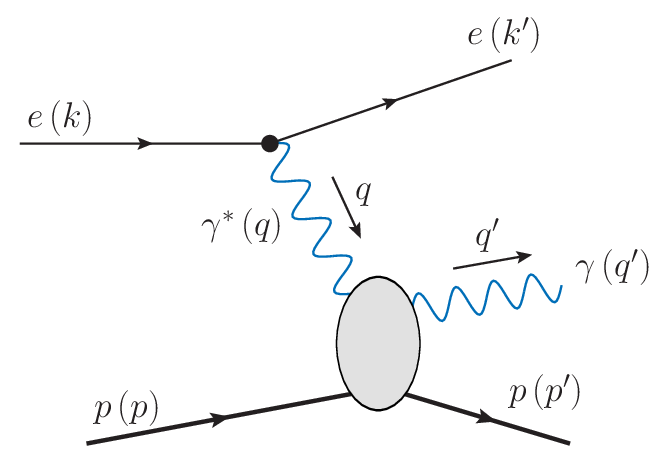}
\caption{
DVCS contribution to $ep \to ep \gamma$ (\ref{2.100}).}
\label{fig:DVCS}
\end{figure}

For an initial virtual photon $\gamma^{*}$ the reaction (\ref{2.1})
is extracted from $ep \to ep \gamma$ scattering
(see Fig.~\ref{fig:DVCS})
\begin{equation}
e(k) + p (p,\lambda) \to 
e(k') + \gamma(q', \epsilon') + p (p',\lambda')\,.
\label{2.100}
\end{equation}
%
Here the Bethe-Heitler process and DVCS contribute with
the latter corresponding to electroproduction of the $\gamma p$ state.
A detailed kinematic analysis of electroproduction reactions
can e.g. be found in \cite{Arens:1996xw} where, in particular,
many relations of variables in the rest system 
of the original proton and in the system used in HERA
experiments are given.
The standard kinematic variables for 
(\ref{2.1}) and (\ref{2.100}), setting the electron mass to zero,
are (see Table~1 of \cite{Arens:1996xw})
%
\begin{eqnarray}
&& q = k - k'\,, \quad q^{2} = -Q^{2}\,, \nonumber \\
&& s = (p + k)^{2}\,, \nonumber \\
&& t = (p - p')^{2}\,, \quad t \leqslant -|t|_{\rm min}\,,\nonumber \\
&& W^{2} = (p + q)^{2}\,, \nonumber \\
&& x = \frac{Q^{2}}{2 p \cdot q}
     = \frac{Q^{2}}{W^{2} + Q^{2} - m_{p}^{2}} \,, \nonumber \\
&& y = \frac{p \cdot q}{p \cdot k}
     = \frac{W^{2} + Q^{2} - m_{p}^{2}}{s - m_{p}^{2}}\,,\nonumber \\
&& \varepsilon = \frac{2 (1 - y) - 2 xy m_{p}^{2} (s - m_{p}^{2})^{-1}}
                    {1 + (1 - y)^{2} + 2 xy m_{p}^{2} (s - m_{p}^{2})^{-1}}\,.
\label{2.101}
\end{eqnarray}
Here $\varepsilon$ is the ratio of longitudinal and transverse
polarization strengths of the virtual photon $\gamma^{*}$
and $t = -|t|_{\rm min}$ corresponds to forward scattering.
For $Q^{2} = 0$ we have $|t|_{\rm min} = 0$
and for $W^{2} \gg Q^{2}, m_{p}^{2}$
we find (see also Appendix~A of \cite{Arens:1996xw})
\begin{equation}
|t|_{\rm min} = Q^{2} \left[ \frac{Q^{2} m_{p}^{2}}{W^{4}} + 
{\cal O}\left(\frac{Q^{6}}{W^{6}}, \frac{Q^{4} m_{p}^{2}}{W^{6}}, \frac{Q^{2} m_{p}^{4}}{W^{6}},\frac{m_{p}^{6}}{W^{6}} \right) \right] \,.
\label{2.101a}
\end{equation}

We assume for (\ref{2.100}) unpolarized initial particles
and no observation of the polarization of final state particles.

Adapting (3.20) and (3.21) of \cite{Arens:1996xw} 
to the reaction (\ref{2.100})
and integrating over the azimuthal angle $\varphi$
defined in (2.1) of \cite{Arens:1996xw}
we get for the DVCS part of (\ref{2.100})
\begin{eqnarray}
\frac{d\sigma (ep \to ep \gamma)}{dydQ^{2}dt}
= \Gamma_{e p \gamma} 
\left( 
\frac{d\sigma_{\rm T}}{dt}(Q^{2},W^{2},t) 
+ \varepsilon \frac{d\sigma_{\rm L}}{dt}(Q^{2},W^{2},t)
\right)\,.
\label{2.102}
\end{eqnarray}
Here, with Hand's convention \cite{Hand:1963bb},
the $\gamma^{*}$ flux factor reads
\begin{eqnarray}
\Gamma_{e p \gamma} =
\frac{\alpha_{\rm em}}{\pi y Q^{2}}
\left( 1 - y + \frac{1}{2}y^{2} 
+ xy \frac{m_{p}^{2}}{s - m_{p}^{2}}\right)
\frac{W^{2} - m_{p}^{2}}{W^{2} - m_{p}^{2} + Q^{2} 
+ 4 x m_{p}^{2}}
\,,
\label{2.103}
\end{eqnarray}
where $\alpha_{\rm em}$ is the fine structure constant.
Furthermore, $d\sigma_{\rm T}/dt$ and 
$d\sigma_{\rm L}/dt$ are the differential cross sections
for $\gamma^{*} p \to \gamma p$ for transverse and longitudinal
polarization of the~$\gamma^{*}$,
\begin{eqnarray}
&&\frac{d\sigma_{\rm T}}{dt}(Q^{2},W^{2},t) =
\frac{1}{2} \left( \frac{d\sigma_{++}}{dt}(Q^{2},W^{2},t) 
                 + \frac{d\sigma_{--}}{dt}(Q^{2},W^{2},t) \right)
                 \,, \nonumber\\
&&\frac{d\sigma_{\rm L}}{dt}(Q^{2},W^{2},t) = 
\frac{d\sigma_{00}}{dt}(Q^{2},W^{2},t)\,,
\label{2.104}
\end{eqnarray}
where
\begin{eqnarray}
\frac{d\sigma_{mm}}{dt}(Q^{2},W^{2},t)
&=& \frac{1}{16 \pi (W^{2} - m_{p}^{2}) 
\sqrt{(W^{2} - m_{p}^{2} + Q^{2})^{2} + 4 m_{p}^{2} Q^{2}}} \nonumber\\
&& \times 
\frac{1}{2}\sum_{\lambda, \lambda', a}
\big\vert \Braket{\gamma(q', \epsilon'_{a}),p (p',\lambda'), {\rm out}
|e J_{\mu}(0) \epsilon_{m}^{\mu}| p (p,\lambda)} \big\vert^{2} 
\,.
\label{2.105}
\end{eqnarray}
Here $e J_{\mu}$ is the electromagnetic current,
$\epsilon_{m} (m = \pm, 0)$ are the standard $\gamma^{*}$
polarization vectors for right and left circular
and longitudinal polarization
[see (3.11)--(3.14) of \cite{Arens:1996xw}]
and $\epsilon'_{a} (a = 1, 2)$
are the polarization vectors of the real photon in the final state.

So far, everything is completely general. 
Now we consider the small $x$ region of DVCS:
\begin{eqnarray}
W \geqslant m_{p}, \sqrt{|t|}\,, \quad 
|t| \lesssim 1 \; {\rm GeV}^{2}\,, \quad 
x \leqslant 0.02\,.
\label{2.106}
\end{eqnarray}
There we describe the amplitude for (\ref{2.1})
in terms of the exchanges of soft ($\Pom_{1}$)
and hard ($\Pom_{0}$) pomeron and the reggeons
$f_{2 \Reg}$ and $a_{2 \Reg}$.
Only $C = +1$ exchanges can contribute in (\ref{2.1}).
The properties of $\Pom_{0}, \Pom_{1}, f_{2 \Reg}$, 
and $a_{2 \Reg}$, and their couplings to protons and photons
will be taken as in \cite{Britzger:2019lvc}.
In this way we get
%
\begin{eqnarray}
&&\Braket{
\gamma(q', \epsilon'),p (p',\lambda'), {\rm out}
|e J_{\nu}(0) \epsilon^{\nu}| p (p,\lambda)} 
\equiv
(\epsilon'^{\mu})^* {\cal M}_{\mu \nu, \lambda' \lambda}\, \epsilon^{\nu}
\nonumber \\
&& 
=-(\epsilon'^{\mu})^*  
\sum_{j = 0, 1}
\Gamma_{\mu \nu \kappa \rho}^{(\Pom_{j} \gamma^{*} \gamma^{*})}(q',q)\, \epsilon^{\nu}\,
\Delta^{(\Pom_{j})\,\kappa \rho, \alpha\beta}(W^{2},t)\, 
\bar{u}_{\lambda'}(p') 
\Gamma_{\alpha\beta}^{(\Pom_{j} pp)}(p',p)
u_{\lambda}(p) 
\nonumber \\  
&& \quad + (\Pom_{j} \to f_{2 \Reg}, a_{2 \Reg})\,.
\label{2.3}
\end{eqnarray}
Here $\Delta^{(\Pom_{j})}$ and $\Gamma^{(\Pom_{j} pp)}$ 
denote the effective propagator and proton vertex function,
respectively, for the tensor pomeron $\Pom_{j}$.
The corresponding expressions, 
as given in 
Appendix~A of \cite{Britzger:2019lvc}, are as follows
\begin{eqnarray}
&&i\Delta^{(\Pom_{j})}_{\mu \nu, \kappa \lambda}(W^{2},t) = 
\frac{1}{4W^{2}} \left( g_{\mu \kappa} g_{\nu \lambda} 
                  + g_{\mu \lambda} g_{\nu \kappa}
                  - \frac{1}{2} g_{\mu \nu} g_{\kappa \lambda} \right)
(-i W^{2} \tilde{\alpha}'_{j})^{\alpha_{j}(t)-1}\,,
\label{2.31}\\
&&i\Gamma_{\mu \nu}^{(\Pom_{j} pp)}(p',p)
=-i 3 \beta_{jpp} F_{1}^{(j)}(t)
\left\lbrace 
\frac{1}{2} 
\left[ \gamma_{\mu}(p'+p)_{\nu} 
     + \gamma_{\nu}(p'+p)_{\mu} \right]
- \frac{1}{4} g_{\mu \nu} (\slash{p}' + \slash{p})
\right\rbrace. \nonumber \\
\label{2.32}
\end{eqnarray}
The coupling constants $\beta_{j pp}$ of the pomerons
to protons are
$\beta_{0 pp} = \beta_{1 pp} = 1.87$~GeV$^{-1}$.
The \textit{Ans{\"a}tze} 
for effective propagators and vertices
for the $f_{2 \Reg}$ and $a_{2 \Reg}$ reggeons
are analogous to (\ref{2.31}) and (\ref{2.32}),
respectively.
We shall assume identical trajectories for $f_{2 \Reg}$ 
and $a_{2 \Reg}$ and combine their contribution in (\ref{2.3})
to a reggeon $\Reg_{+}$ term labeled $j = 2$
as explained in Appendix~A of \cite{Britzger:2019lvc};
see (A20)--(A31) of \cite{Britzger:2019lvc}.

The pomeron and reggeon trajectory functions
are assumed to be of linear form
%
\begin{eqnarray}
\alpha_{j}(t) = \alpha_{j}(0)+\alpha'_{j}\,t\,,
\quad
\alpha_{j}(0) = 1 + \epsilon_{j}\,,
\quad
(j = 0, 1, 2)\,.
\label{trajectories}
\end{eqnarray}
%
In (\ref{2.31}) the parameters multiplying the squared energy
$W^{2}$ 
in the effective propagators
are taken as
$\tilde{\alpha}'_{j} = \alpha'_{j}$.
The slope parameters $\alpha'_{j}$ 
are taken as the default values from 
\cite{Britzger:2019lvc}:
$\alpha'_{1} = \alpha'_{0} = 0.25\;{\rm GeV}^{-2}$,
$\alpha'_{2} = 0.9\;{\rm GeV}^{-2}$.
The intercept parameters of the trajectories (\ref{trajectories})
were determined from detailed comparison
of the two-tensor-pomeron model with the DIS HERA data
and photoproduction data in Ref.~\cite{Britzger:2019lvc}:
\begin{eqnarray}
{\rm soft \; pomeron} \;\Pom_{1}:\quad &&\epsilon_{1} = 0.0935(^{+76}_{-64})\,,
\label{FIT_parameters_P1}
\\
{\rm hard \; pomeron} \;\Pom_{0}:\quad &&\epsilon_{0} = 0.3008(^{+73}_{-84})\,,
\label{FIT_parameters_P0}\\
{\rm reggeon} \;\Reg_{+}:\quad &&\alpha_{2}(0) = 0.485(^{+88}_{-90})\,.
\label{FIT_parameters_R}
\end{eqnarray}
%
Note that for the fit to the DIS structure functions
and the total photoabsorption cross section,
being related to the absorptive parts of the \textit{forward}
virtual and real Compton amplitudes,
only the values of the \textit{intercepts} of the pomerons
and reggeons enter;
see Eqs. (3.7)--(3.10) of \cite{Britzger:2019lvc}.
What also enters there are the scale or $W^{2}$ parameters
$\tilde{\alpha}'_{j}$ ($j = 0, 1, 2$).
For convenience these were chosen equal to the slope parameters
$\alpha'_{j}$ ($j = 0, 1, 2$); see (A3), (A7), 
and (A22) of \cite{Britzger:2019lvc}.
But from (3.7)--(3.10) of \cite{Britzger:2019lvc} we see that
a change of the $\tilde{\alpha}'_{j}$ parameters affects only 
the values of the coupling parameters and does not affect
the values of the intercepts.
In (\ref{FIT_parameters_P1})--(\ref{FIT_parameters_R}) above
we quote the values of 
$\epsilon_{j} = \alpha_{j}(0)-1$ for $j = 0, 1$
and $\alpha_{2}(0)$ with errors as they were obtained
in \cite{Britzger:2019lvc}.
The issue of linearity or non-linearity of the Regge trajectories
plays absolutely no role there.
In our present work, however, the linearity assumption
(\ref{trajectories}) \textit{is important}.
For a discussion of the linearity of the trajectories see e.g.
the reviews \cite{Collins:1977,Donnachie:2002en}.
Non-linear trajectories are discussed, for instance, in
\cite{Brisudova:1999ut,Tang:2000tb,Fiore:2000fp,Boschi-Filho:2005xct,
Godizov:2007dy,Fiore:2008tp,
Jenkovszky:2017efs,Jenkovszky:2018itd,Szanyi:2019kkn}.

The \textit{Ans{\"a}tze} for the $\Pom \gamma^{*} \gamma^{*}$ and
$\Reg_{+}\gamma^{*} \gamma^{*}$ 
coupling functions for both real and virtual photons
are discussed in \cite{Britzger:2019lvc}.
The $\Pom_{j} \gamma^{*} \gamma^{*}$ vertex,
coupling the soft ($j = 0$) and hard ($j = 1$) pomeron 
to two (virtual or real) photons, reads 
[see (A18) of \cite{Britzger:2019lvc}]
\begin{eqnarray}
&&i\Gamma^{(\Pom_{j} \gamma^{*} \gamma^{*})}_{\mu \nu \kappa \rho}(q',q) =
i\left[
2a_{j \gamma^{*} \gamma^{*}}(q^{2},q'^{2},t)\,\Gamma^{(0)}_{\mu \nu \kappa \rho}(q',-q)\,
     - b_{j \gamma^{*} \gamma^{*}}(q^{2},q'^{2},t)\,\Gamma^{(2)}_{\mu \nu \kappa \rho}(q',-q)
     \right], \nonumber \\
&&{\rm where} \; t = (q - q')^{2}, \quad j = 0,1\,.
\label{2.5}
\end{eqnarray}  
The coupling functions $a_{j \gamma^{*} \gamma^{*}}$ 
and $b_{j \gamma^{*} \gamma^{*}}$ have dimensions 
GeV$^{-3}$ and GeV$^{-1}$, respectively.
The rank-4 tensor functions 
$\Gamma^{(i)}_{\mu \nu \kappa \rho}$ ($i = 0, 2$)
are defined in (A13)--(A17) of \cite{Britzger:2019lvc}.
The $\Reg_{+} \gamma^{*} \gamma^{*}$ vertex
for real and virtual photons has the same structure 
as (\ref{2.5}) and is labeled with $j = 2$;
see (A27)--(A31) of \cite{Britzger:2019lvc}.

Inserting (\ref{2.31}), (\ref{2.32}) and (\ref{2.5})
in (\ref{2.3}) we get
\begin{eqnarray}
{\cal M}_{\mu \nu, \lambda' \lambda}
&=&-i \sum_{j = 0, 1, 2}
\left[ 2a_{j \gamma^{*} \gamma^{*}}(q^{2},0,t)\,\Gamma^{(0)}_{\mu \nu \kappa \rho}(q',-q)\,
- b_{j \gamma^{*} \gamma^{*}}(q^{2},0,t)\,\Gamma^{(2)}_{\mu \nu \kappa \rho}(q',-q) \right] \nonumber \\
&&\times \frac{1}{2W^{2}} 
(-i W^{2} \alpha'_{j})^{\alpha_{j}(t)-1} \;
3 \beta_{jpp} F_{1}^{(j)}(t)\,
\bar{u}_{\lambda'}(p') 
\gamma^{\kappa}(p' + p)^{\rho}
u_{\lambda}(p) \,.
\label{2.200}
\end{eqnarray}
From (\ref{2.200}) we can read off the parameters of our model.
Taking the pomeron and reggeon parameters as specified above
for granted ${\cal M}_{\mu \nu, \lambda' \lambda}$
is parametrized by the functions
\begin{eqnarray}
&&a_{j \gamma^{*} \gamma^{*}}(q^{2},0,t)\,
\beta_{jpp} F_{1}^{(j)}(t)\,,
\nonumber \\
&&b_{j \gamma^{*} \gamma^{*}}(q^{2},0,t)\,
\beta_{jpp} F_{1}^{(j)}(t)\,,
\nonumber \\
&&j = 0, 1, 2\,.
\label{2.201}
\end{eqnarray}
Here we consider the $\beta_{jpp}$ as normalization constants
with values
$\beta_{0 pp} = \beta_{1 pp} = 1.87$~GeV$^{-1}$,
$\beta_{2 pp} = 3.68$~GeV$^{-1}$,
as given in Appendix~A of \cite{Britzger:2019lvc}.

For forward virtual Compton scattering,
$q^{2} = q'^{2} = -Q^{2}$ and $t = 0$, 
we have from (A19), (A30), and (A31) of \cite{Britzger:2019lvc}
\begin{eqnarray}
&&a_{j \gamma^{*} \gamma^{*}}(-Q^{2},-Q^{2},0) = e^{2} \hat{a}_{j}(Q^{2})\,,
\nonumber \\
&&b_{j \gamma^{*} \gamma^{*}}(-Q^{2},-Q^{2},0) = e^{2} \hat{b}_{j}(Q^{2}) \,, 
\nonumber \\
&&j = 0, 1, 2\,.
\label{2.7_DIS}
\end{eqnarray}  
The coupling functions 
$\hat{a}_{j}(Q^{2})$ and $\hat{b}_{j}(Q^{2})$
were determined from the global fit to HERA inclusive DIS data
for $Q^{2} < 50$~GeV$^{2}$ and $x < 0.01$
and the ($Q^{2} = 0$) photoproduction data;
see Appendix~C and Table~IV of~\cite{Britzger:2019lvc}.


The $Q^{2}$ and $t$ dependencies of the coupling functions
in the $\Pom_{j} \gamma^{*}\gamma^{*}$ 
and $\Reg_{+} \gamma^{*}\gamma^{*}$ vertices
in (\ref{2.200}) 
must be determined from a comparison to experimental data.
As we shall show in Sec.~\ref{sec:results} we obtain quite
satisfactory descriptions of 
the DVCS HERA and elastic $\gamma p$-scattering FNAL data
with the following assumptions:
\begin{eqnarray}
&&a_{j \gamma^{*} \gamma^{*}}(q^{2},0,t) = 
e^{2} \hat{a}_{j}(Q^{2}) F^{(j)}(t)\,, \quad (j = 0, 1, 2)\,,
\nonumber \\
&&b_{j \gamma^{*} \gamma^{*}}(q^{2},0,t) = 
e^{2} \hat{b}_{j}(Q^{2}) F^{(j)}(t)\,, \quad (j = 2)\,,
\label{2.7}
\end{eqnarray}  
and for alternative fits 1 and 2
%
\begin{eqnarray}
{\rm FIT~1}: \;
&&b_{1 \gamma^{*} \gamma^{*}}(q^{2},0,t) = e^{2} \hat{b}_{1}(0) 
(1 + Q^{2}/\Lambda_{1}^{2})^{-1.2}\, F^{(1)}(t)\,,
\quad \Lambda_{1} = 1.4\;{\rm GeV}\,, \nonumber\\
&&b_{0 \gamma^{*} \gamma^{*}}(q^{2},0,t) = e^{2} \hat{b}_{0}(Q^{2}) 
F^{(0)}(t)\,,
\label{FIT1} \\
{\rm FIT~2}:\;
&&b_{1 \gamma^{*} \gamma^{*}}(q^{2},0,t) = e^{2} \hat{b}_{1}(0) 
(1 + Q^{2}/\Lambda_{2}^{2})^{-2.0} \, F^{(1)}(t)\,,
\quad \Lambda_{2} = 2.0\;{\rm GeV}\,, \nonumber \\
&&
b_{0 \gamma^{*} \gamma^{*}}(q^{2},0,t) =
\left\{ \begin{array}{lr} 
e^{2} \hat{b}_{0}(Q^{2}) \, F^{(0)}(t)
& {\rm for}\ Q^{2} < 1.5~{\rm GeV}^{2} \\ 
e^{2} \Lambda_{0} (1 + Q^{2}/\Lambda_{3}^{2})^{-0.6} \, F^{(0)}(t)
& {\rm for}\ Q^{2} \geq 1.5~{\rm GeV}^{2} 
\end{array}\right. \,, \nonumber \\
&& \Lambda_{0} = 9.46 \times 10^{-3} \; {\rm GeV}^{-1}\,, \quad 
\Lambda_{3} = 2.3\;{\rm GeV}\,.
\label{FIT2}
\end{eqnarray}

\begin{figure}[!ht]
\includegraphics[width=0.8\textwidth]{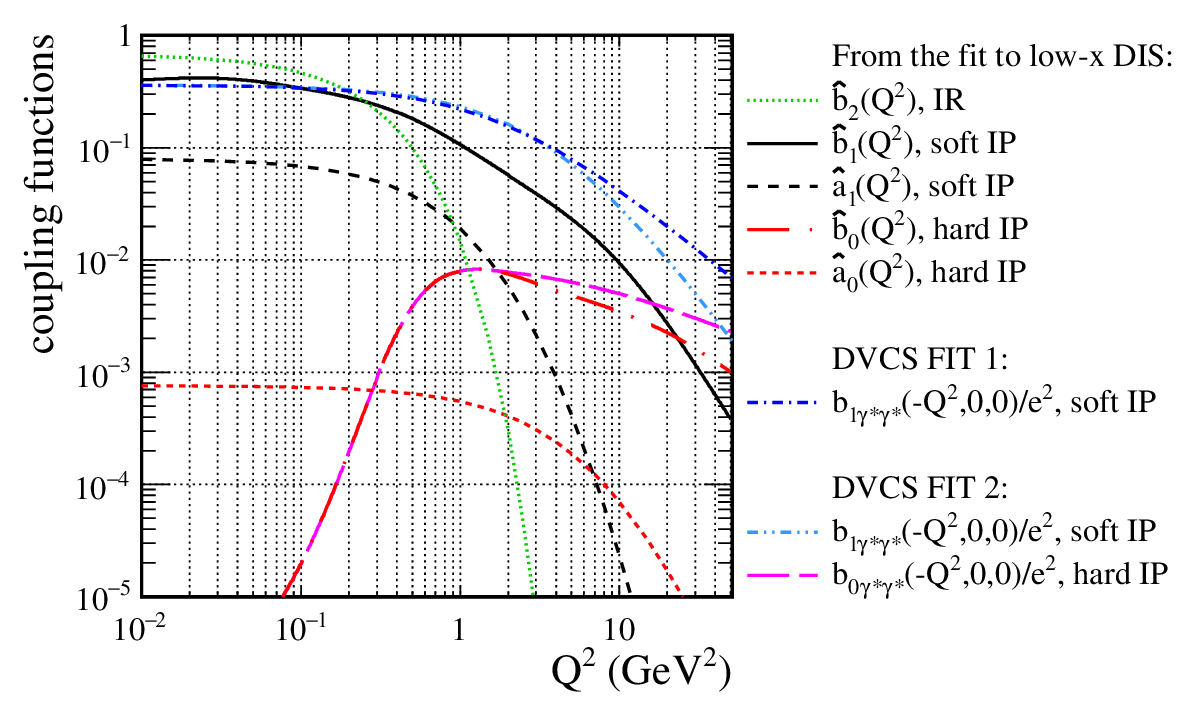}
\caption{\label{fig:FIT}
\small
The pomeron- and reggeon-$\gamma^{*}\gamma^{*}$ coupling functions
$\hat{a}_{j}(Q^{2})$ and $\hat{b}_{j}(Q^{2})$
from (\ref{2.7_DIS}),
for $\Reg_{+}$ reggeon ($j = 2$, green dotted line),
soft pomeron ($j = 1$, black solid and dashed lines),
and hard pomeron 
($j = 0$, red long-dash-dotted and short-dashed lines),
as determined in \cite{Britzger:2019lvc}.
The experimental uncertainties of the fit are not shown here.
For this we refer to Figs.~12--15 of \cite{Britzger:2019lvc}.
The blue dash-dotted line
corresponds to the $b_{1\gamma^{*}\gamma^{*}}(-Q^{2},0,0)/e^{2}$ coupling function
used in the FIT~1 (\ref{FIT1}).
The azure dash-dot-dotted line 
and the magenta long-dashed line are for
$b_{1\gamma^{*}\gamma^{*}}(-Q^{2},0,0)/e^{2}$ and $b_{0\gamma^{*}\gamma^{*}}(-Q^{2},0,0)/e^{2}$,
respectively,
in the FIT~2 (\ref{FIT2}).
The $a$ and $b$ parameters are plotted in units 
GeV$^{-3}$ and GeV$^{-1}$, respectively.}
\end{figure}

The coupling functions as a function of $Q^{2}$
used in the calculations are shown in Fig.~\ref{fig:FIT}.
Shown are
only the middle values of the
pomeron- and reggeon-$\gamma^{*}\gamma^{*}$ coupling functions 
up to $Q^{2} = 50$~GeV$^{2}$ 
determined from a global fit in \cite{Britzger:2019lvc}.
For a detailed discussion of the fit quality see \cite{Britzger:2019lvc}.
According to \cite{Britzger:2019lvc},
for the $\Reg_{+}$-reggeon term,
the function $\hat{a}_{2}(Q^{2})$ is assumed to be zero
while the function $\hat{b}_{2}(Q^{2})$
(see the green dotted line)
vanishes rapidly with increasing $Q^{2}$.
We see that the soft pomeron function
$\hat{b}_{1}$ is larger than the corresponding hard one $\hat{b}_{0}$
up to $Q^{2} \simeq 20$~GeV$^{2}$.
The function $\hat{b}_{0}$
(see the red long-dash-dotted line)
is very small for $Q^{2} \to 0$
and reaches a maximum at $Q^{2} = 1.27$~GeV$^{2}$
with amplitude $\hat{b}_{0} = 0.0082$~GeV$^{-1}$.
This function is important in the large-$Q^{2}$ region.
Thus, it can be expected that the soft pomeron contribution 
will be essential for understanding the HERA DVCS data.
We consider two different parametrizations:
(1) FIT~1 (\ref{FIT1}) with the modification of only one 
pomeron-$\gamma^* \gamma^{*}$ coupling function
$\hat{b}_{1}(Q^{2})$,
corresponding to a ``soft term'';
(2) FIT~2 (\ref{FIT2}) with 
the modification of both $\hat{b}_{1}(Q^{2})$ 
and $\hat{b}_{0}(Q^{2})$. 
Here we increase the hard component
and decrease the soft one relative to FIT~1. 
All other coupling functions,
in both our fits,
are fixed in accord with those from the fit to DIS
of \cite{Britzger:2019lvc}.

We assume, furthermore, that the $t$ dependence
of the $\gamma^{(*)} p \to \gamma p$ amplitudes
is described by the Regge factors and
the following combined functions
\begin{equation}
F_{\rm eff}^{(j)}(t) = 
F^{(j)}(t) \times  F_{1}^{(j)}(t) =
\exp(-b_{j}|t|/2 )\,.
\label{t_dependence_ff}
\end{equation}
Note that the same $t$ dependence is assumed for both $a$ and $b$ 
coupling functions for a given $j$~$(j = 0, 1, 2)$
in Eqs.~(\ref{2.7})--(\ref{FIT2}).
For our study here we assume $b_{1} = b_{2}$.
We take, for both FIT~1 and FIT~2 the same parameters,
$b_{1} = b_{2} = 5.0\; {\rm GeV}^{-2}$
and 
$b_{0} = 1.0\; {\rm GeV}^{-2}$
found by comparison of the theoretical curves
to the experimental data for $d\sigma/dt$.

With this we have specified our model and given the parameter
values for our two fits which we shall compare to the data
in Sec.~\ref{sec:results}.
\section{Results}
\label{sec:results}

In this section we compare
the tensor-pomeron model to the FNAL data 
\cite{Breakstone:1981wm}
on real-photon-proton scattering ($\gamma p \to \gamma p$),
and to HERA data
\cite{ZEUS:2003pwh,H1:2005gdw,ZEUS:2008hcd,H1:2009wnw} 
on DVCS ($\gamma^{*}(Q^{2}) p \to \gamma p$)
for different averaged $W$ and $Q^{2}$.
We shall restrict our discussion to experimental results
that satisfy the conditions $x \approx Q^{2}/W^{2} < 0.02$
and $|t| \lesssim 1$~GeV$^{2}$ where our model should be reliable.
We also have to make a comment on \textit{what} quantity
is measured in \cite{H1:2005gdw}.
For this we compare Eqs.~(6)--(8) of \cite{H1:2005gdw} 
with Eqs.~(\ref{2.102})--(\ref{2.104}).
We cannot see that longitudinally polarized $\gamma^{*}$'s 
do not contribute after integration over azimuthal angles, 
as is claimed in footnote~7 of \cite{H1:2005gdw}.
Thus, to the best of our understanding, the cross section
measured in \cite{H1:2005gdw} must be
\begin{equation}
\frac{d\sigma}{dt} 
= \frac{d\sigma_{\rm T}}{dt} + \varepsilon \frac{d\sigma_{\rm L}}{dt}\,,
\label{3.100}
\end{equation}
where $\varepsilon \approx 1$ for the HERA kinematic region.
In the following we suppose that this is the case.
We shall use the notation
\begin{equation}
\sigma(Q^{2},W^{2})
= \int dt \, \frac{d\sigma}{dt}(Q^{2},W^{2},t)
\label{3.101}
\end{equation}
and similarly for the $\rm T$ and $\rm L$ components.

\begin{figure}[!ht]
\includegraphics[width=0.49\textwidth]{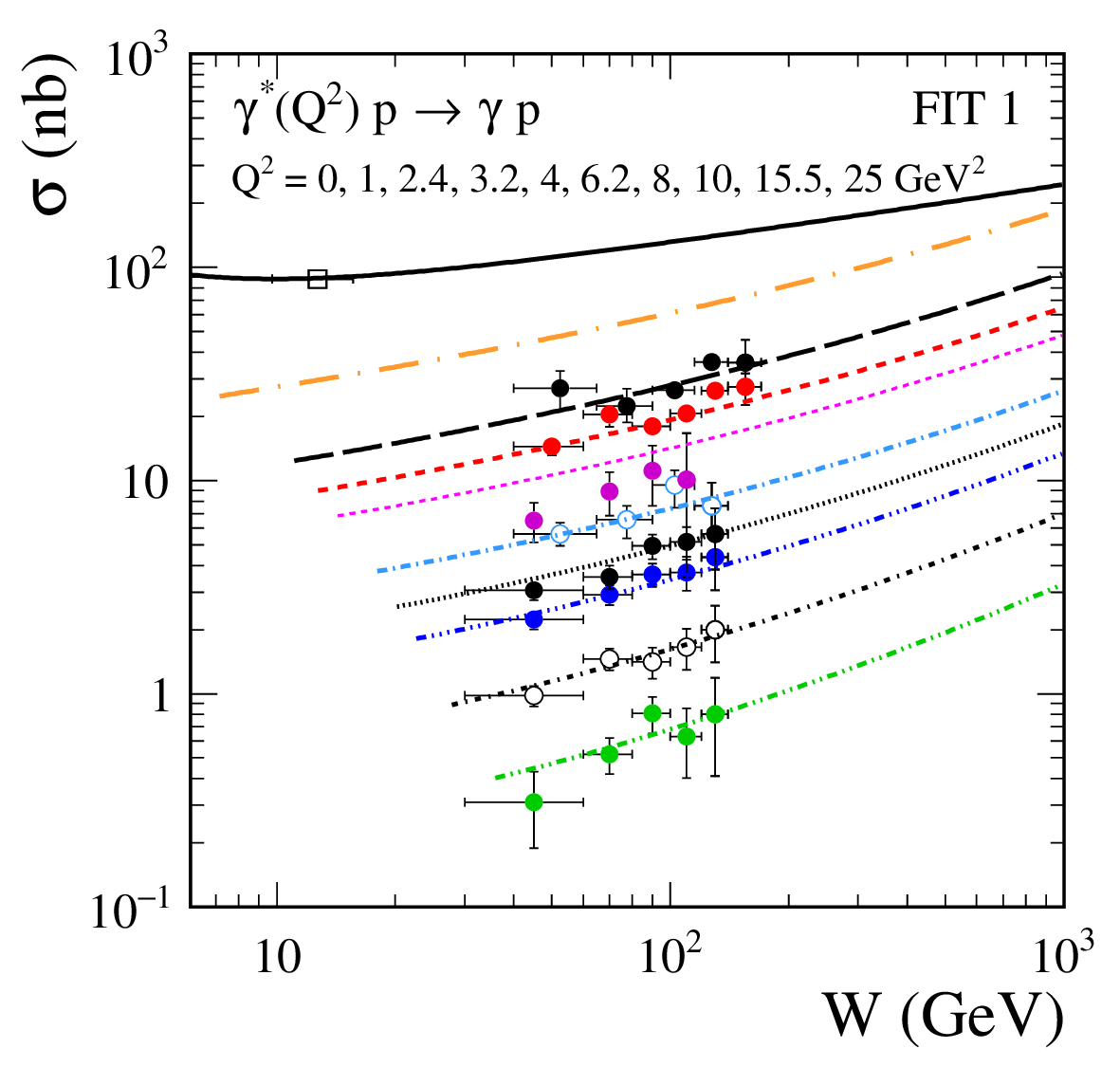}
\includegraphics[width=0.49\textwidth]{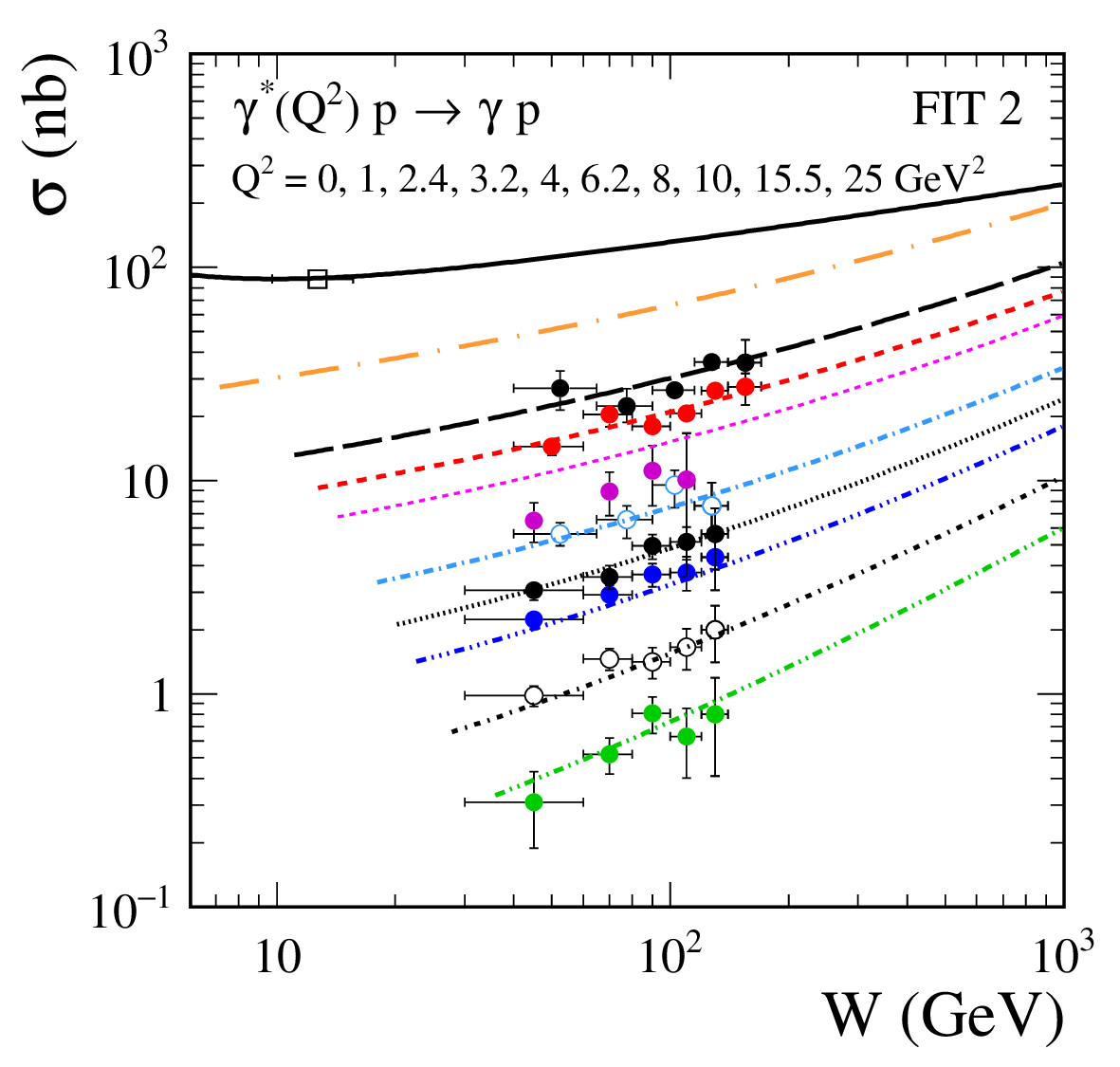}
\includegraphics[width=0.49\textwidth]{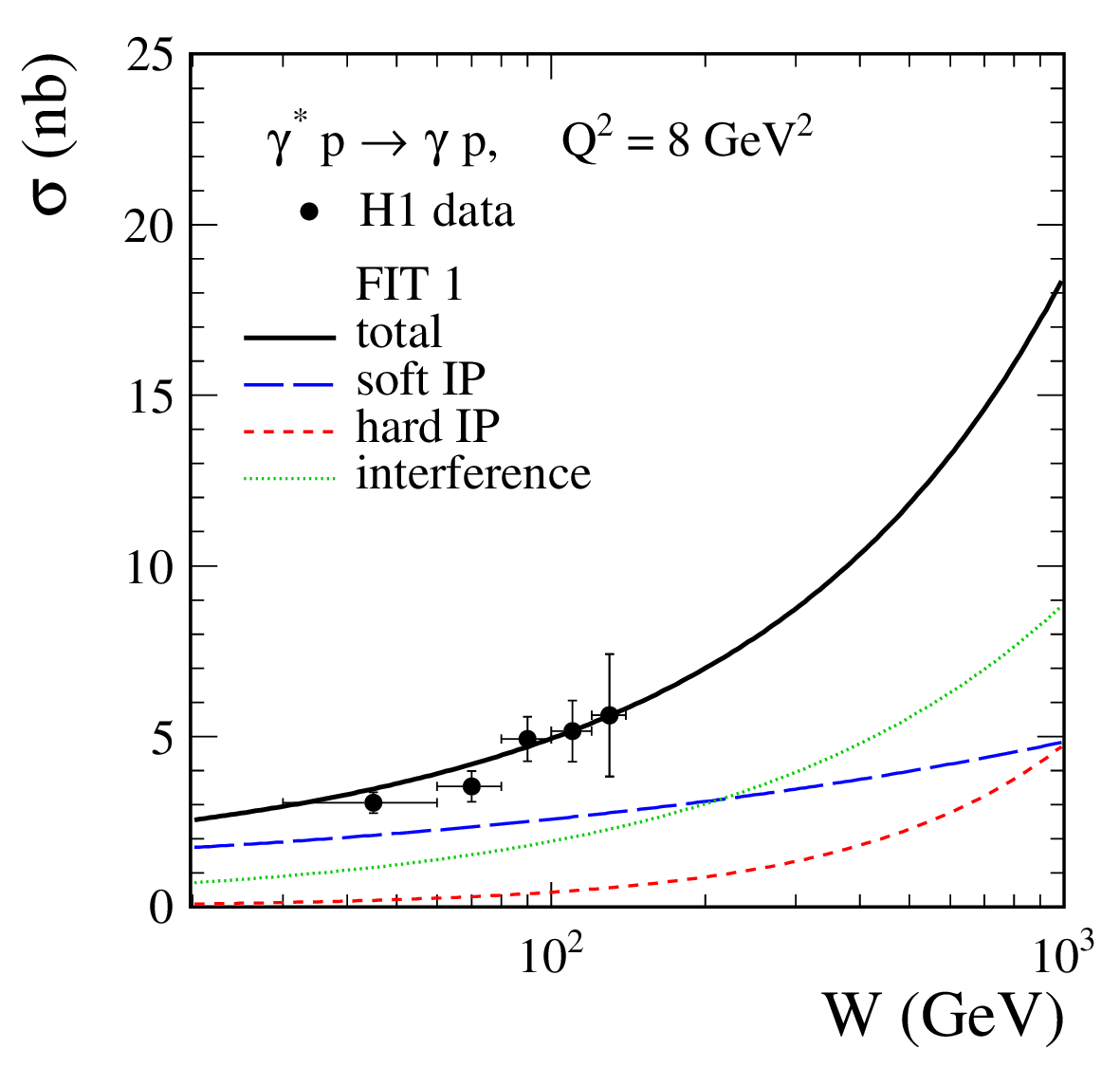}
\includegraphics[width=0.49\textwidth]{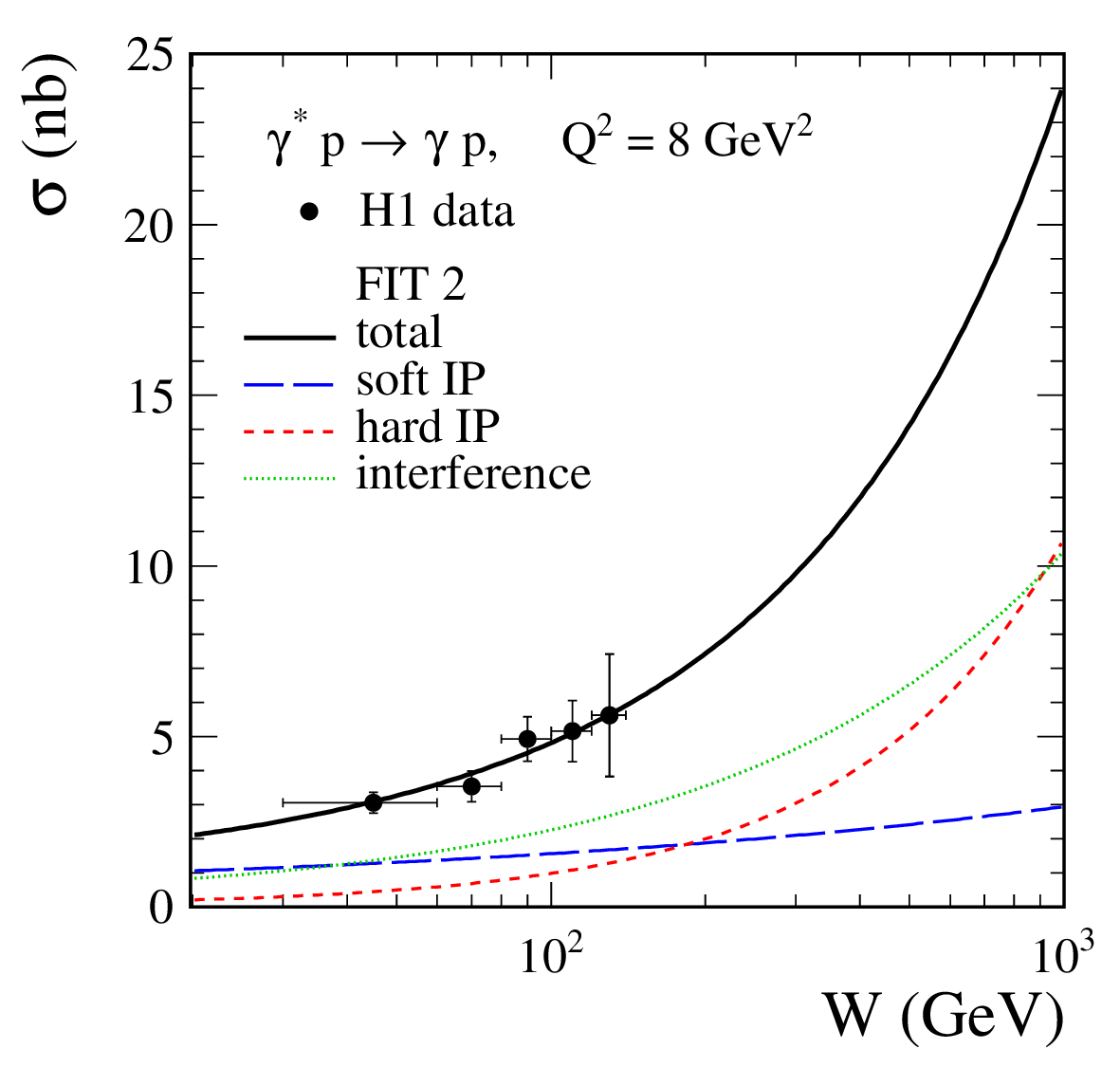}
\caption{\label{fig:1}
\small
Top panels: 
Total cross sections as a function of the c.m. energy $W$
for FIT~1 (left) and FIT~2 (right).
Comparison of theoretical results
to the FNAL data from \cite{Breakstone:1981wm} 
for real Compton scattering ($Q^{2} = 0$)
and to the DVCS HERA data is shown.
The upper black solid line is for $Q^{2} = 0$,
the orange long-dashed-dotted line is for $Q^{2} = 1$~GeV$^{2}$.
The remaining lines correspond to the values
\mbox{$Q^{2} = 2.4, 3.2, 4,
 6.2, 8, 10, 15.5, 25$~GeV$^{2}$}
(from top to bottom) and should be compared with 
the HERA data from
\cite{ZEUS:2003pwh,H1:2005gdw,ZEUS:2008hcd,H1:2009wnw}.
Bottom panels:
Our fit results for $Q^{2} = 8$~GeV$^{2}$
together with the H1 data \cite{H1:2009wnw}.
We show the contributions for soft and hard pomeron separately,
see the blue long-dashed line and the red dashed line,
and their coherent sum (total), see the black solid line.
The interference term is shown separately by the green dotted line.
}
\end{figure}

In Fig.~\ref{fig:1} we show
the $\gamma^{(*)}  p \to \gamma p$ total cross sections
as a function of the c.m. energy $W$.
In the top panels we compare our fits,
FIT~1 (left panel) and FIT~2 (right panel),
to the FNAL result from \cite{Breakstone:1981wm}\footnote{Note
that in \cite{Breakstone:1981wm} the value of 
$\sigma = 88 \pm 4$~nb for the elastic $\gamma p$ cross section
was given. 
This value was obtained by summing 
the beam-energy-averaged differential cross sections 
measured at c.m. energies $W = 9.73-15.645$~GeV
for $|t| \geqslant 0.07$~GeV$^{2}$
with those from the preferred fit~II of \cite{Breakstone:1981wm} 
for $|t| < 0.07$~GeV$^{2}$.} 
where $Q^{2}  = 0$ and to the HERA data from
\cite{ZEUS:2003pwh,H1:2005gdw,ZEUS:2008hcd,H1:2009wnw}
for different values of $Q^{2}$ (the average values).
The complete cross section is a coherent sum of
soft and hard components in the amplitude.
These components have different energy dependence.
For real Compton scattering ($Q^{2} = 0$)
the cross section is dominated by soft-pomeron exchange with
an additional contribution from reggeon exchange at lower energies.
The hard-pomeron contribution is negligibly small there.
The cross sections rise with energy
as $W^{2 \epsilon_{1}}$ at $Q^{2} = 0$
and change to $W^{2 \epsilon_{0}}$ for very high $Q^{2}$.
Here $\epsilon_{1} \approx 0.09$
and $\epsilon_{0} \approx 0.30$
are the intercept parameters
of the soft and hard pomerons, respectively, 
see~(\ref{FIT_parameters_P1}), (\ref{FIT_parameters_P0}).
In our model the dominant contribution 
comes from the $b$-type coupling functions
$b_{j \gamma^{*}\gamma^{*}}$ ($j =0,1$).
Their size, as shown in Fig.~\ref{fig:FIT}, 
differs in the two fits.
We see from the bottom panels of Fig.~\ref{fig:1} that
for higher $Q^{2}$ the soft component slowly decreases
relative to the hard one.
A significant constructive interference effect 
between the soft and hard components is clearly visible.
Here and in the following, the interference term
is calculated as the difference of 
coherent and incoherent cross sections
of the $\Pom_{1}$, $\Pom_{0}$, and $\Reg$ contributions.
The Fits~1 and 2 hardly differ for the $W$ region where
there are data. But for higher $W$ values FIT~2, 
where the contribution from the hard pomeron is enhanced,
gives a steeper rise of the cross section with $W$ 
and especially so for larger $Q^{2}$.
Only future experiments will tell us what happens in nature.

\begin{figure}[!ht]
\includegraphics[width=0.49\textwidth]{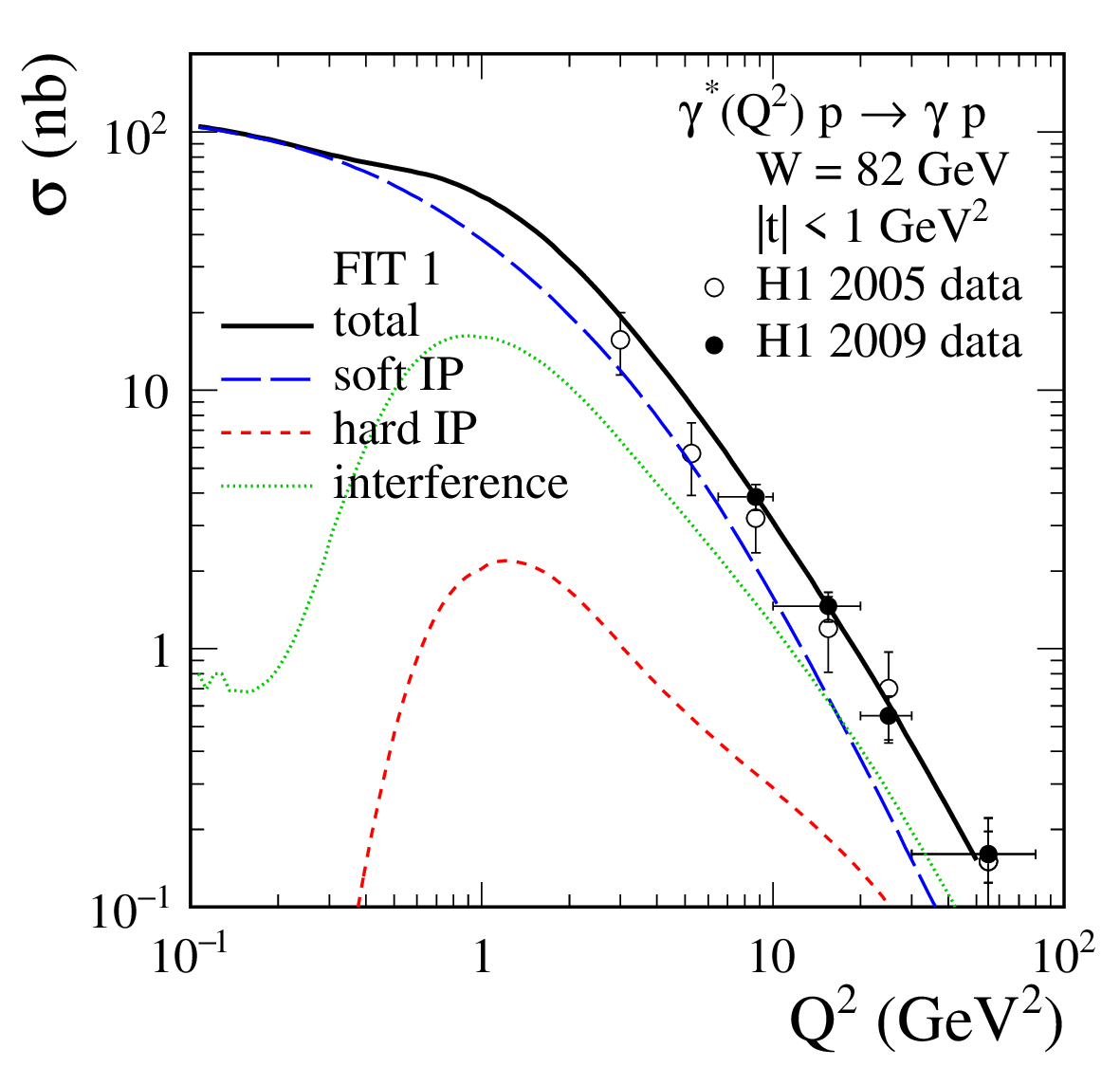}
\includegraphics[width=0.49\textwidth]{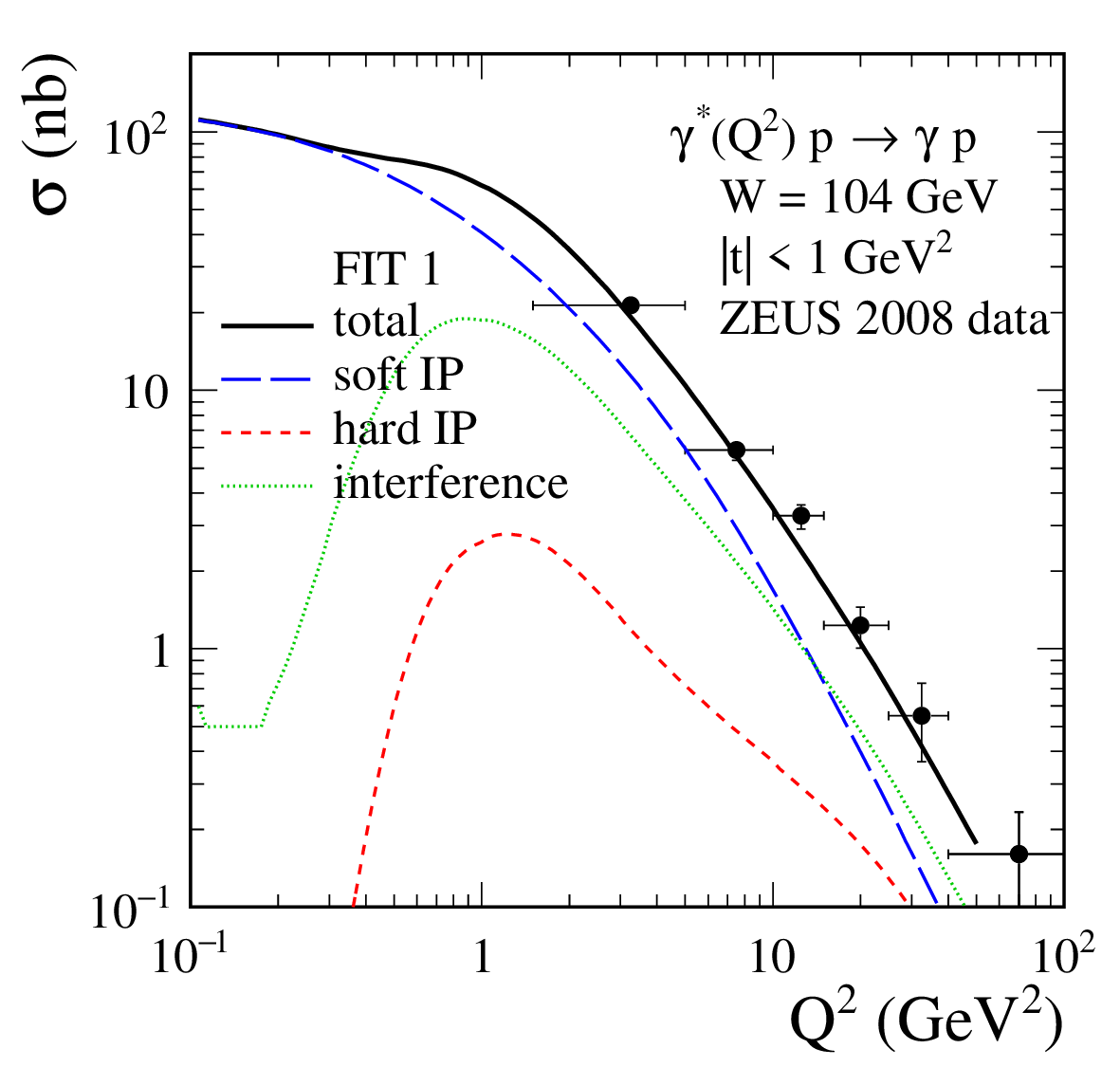}
\includegraphics[width=0.49\textwidth]{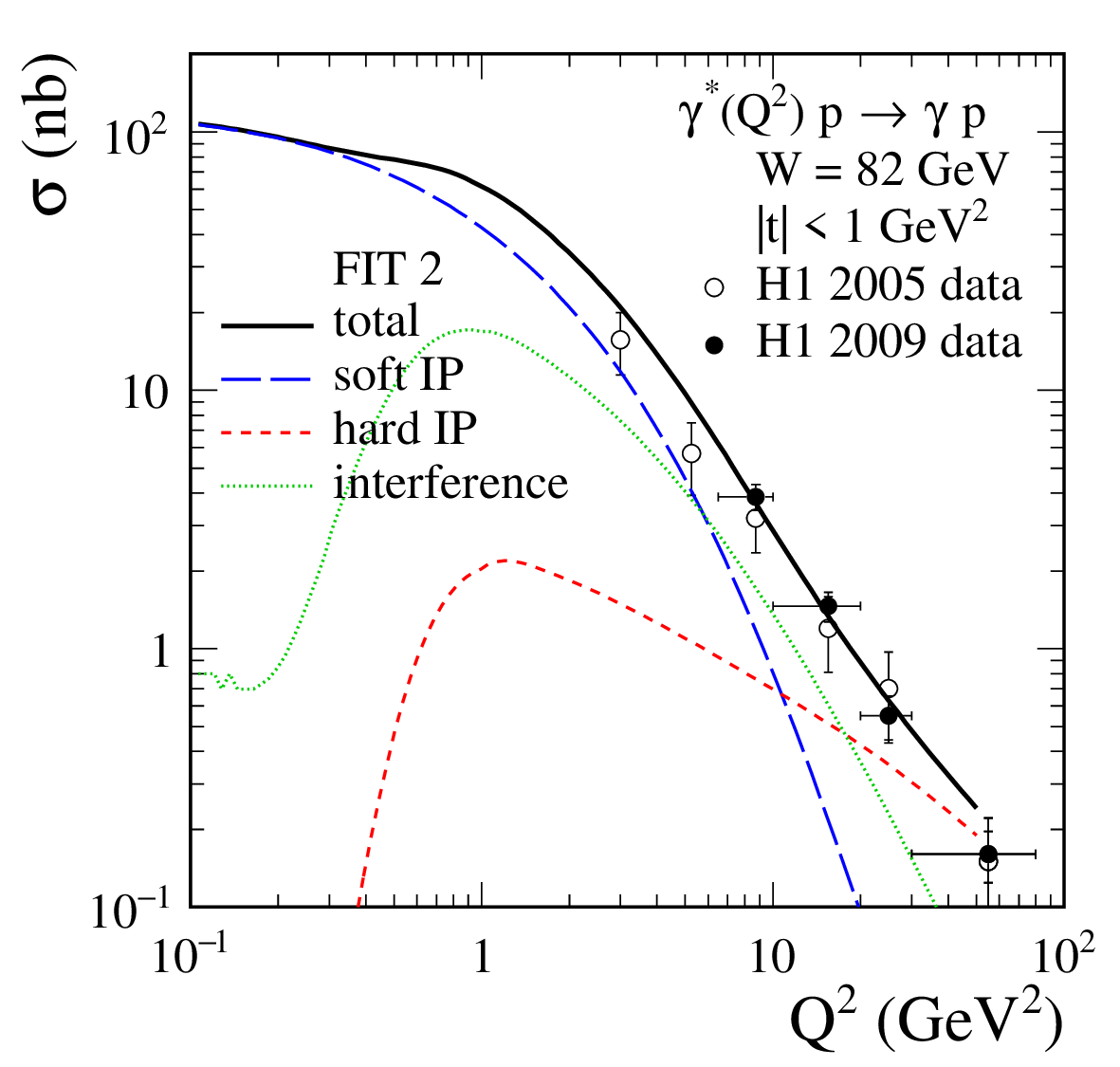}
\includegraphics[width=0.49\textwidth]{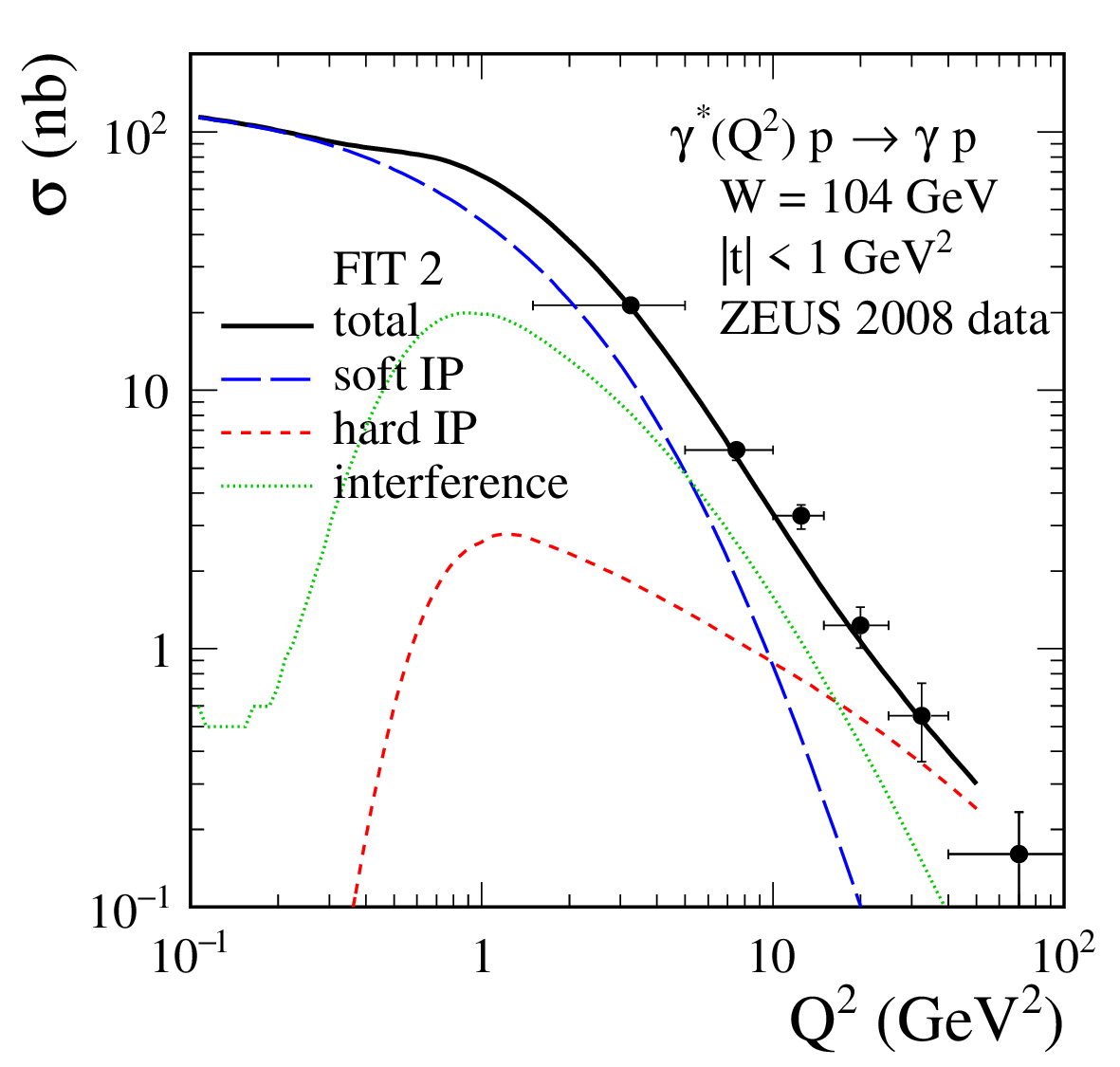}
\caption{\label{fig:2}
\small
Comparison of the cross sections
for the $\gamma^{(*)} (Q^{2}) p \to \gamma p$ reaction
as a function of $Q^{2}$
to the experimental data for $\langle W \rangle = 82$ GeV from 
\cite{H1:2005gdw,H1:2009wnw} (left panels)
and for $\langle W \rangle = 104$ GeV from \cite{ZEUS:2008hcd} (right panels).
Results on top panels correspond to FIT~1 and 
those on the bottom panels to FIT~2.
The meaning of the lines is the same as in the bottom panels
of Fig.~\ref{fig:1}.}
\end{figure}

In Fig.~\ref{fig:2} we show how
the cross section depends on $Q^{2}$
for two c.m. energies $W = 82$ and 104~GeV.
Complete results for our two fits
and the soft and hard components separately
are shown together with the HERA data.
From the top panels (for FIT~1) we see that
the soft pomeron survives to relatively large $Q^2$
and at $Q^2 \simeq 50$~GeV$^{2}$ the interference term
plays an important role in the description of the data.
From the bottom panels (for FIT~2) one can observe that
at $Q^2 \simeq 50$~GeV$^{2}$ the hard pomeron alone
is able to describe the data.
Here the contribution from the interference term
is considerable for intermediate values of $Q^{2} \sim 10$~GeV$^{2}$.
It is interesting to note that 
the interference effect is also large
when the hard component reaches a maximum.
For our fits this is at $Q^{2} \cong 1.3$~GeV$^{2}$.

\begin{figure}[!ht]
\includegraphics[width=0.49\textwidth]{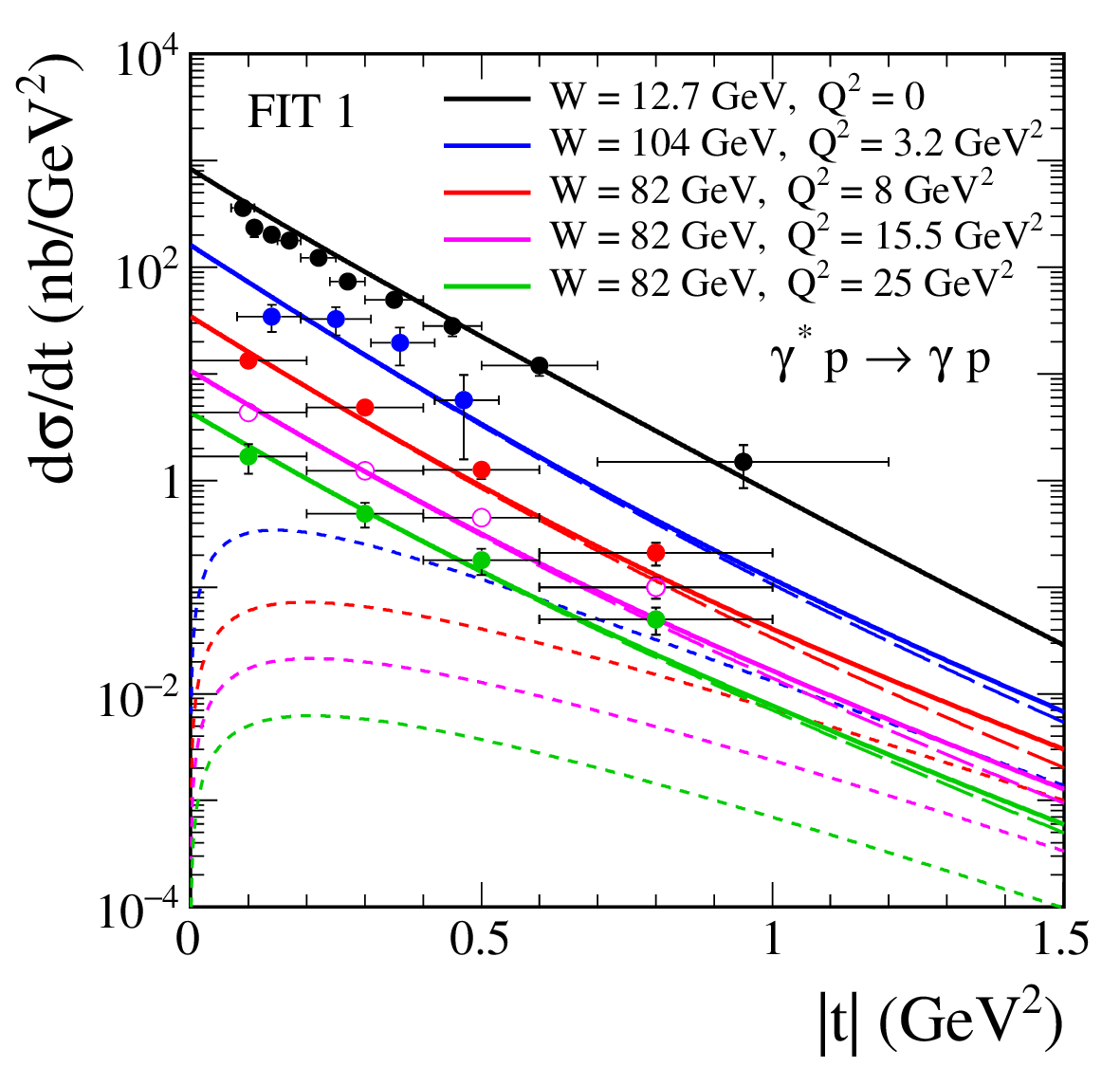}
\includegraphics[width=0.49\textwidth]{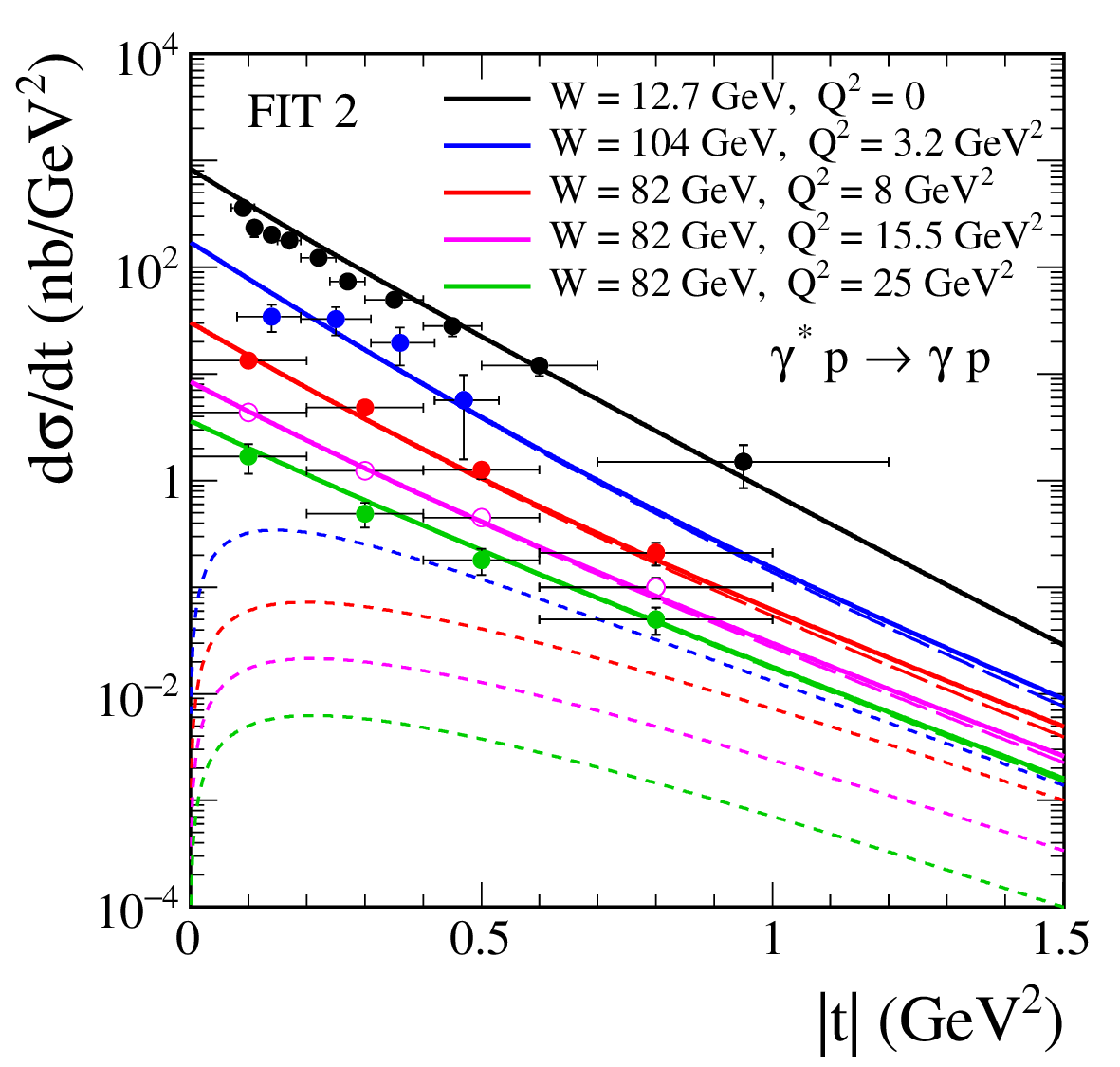}
\includegraphics[width=0.49\textwidth]{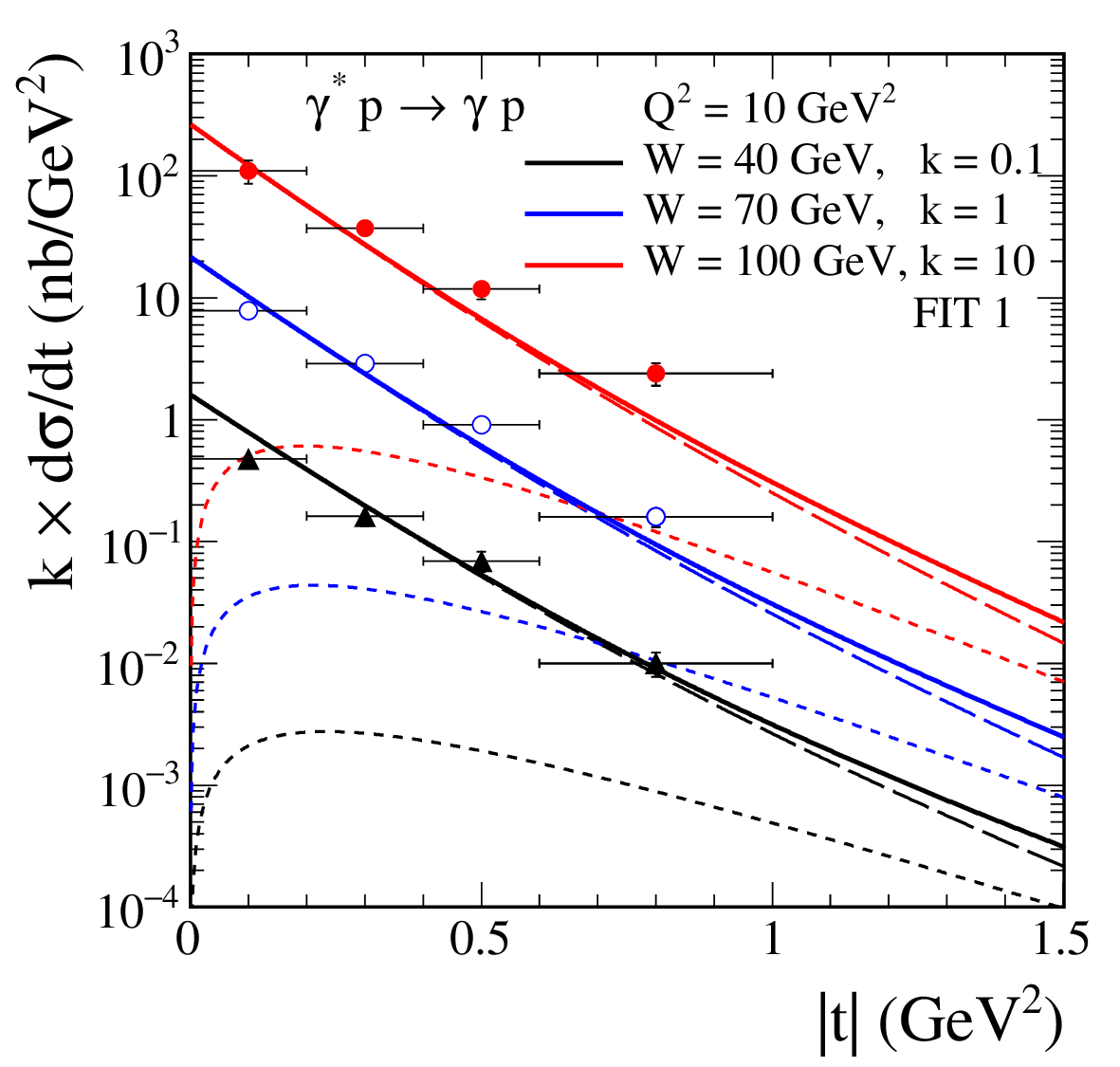}
\includegraphics[width=0.49\textwidth]{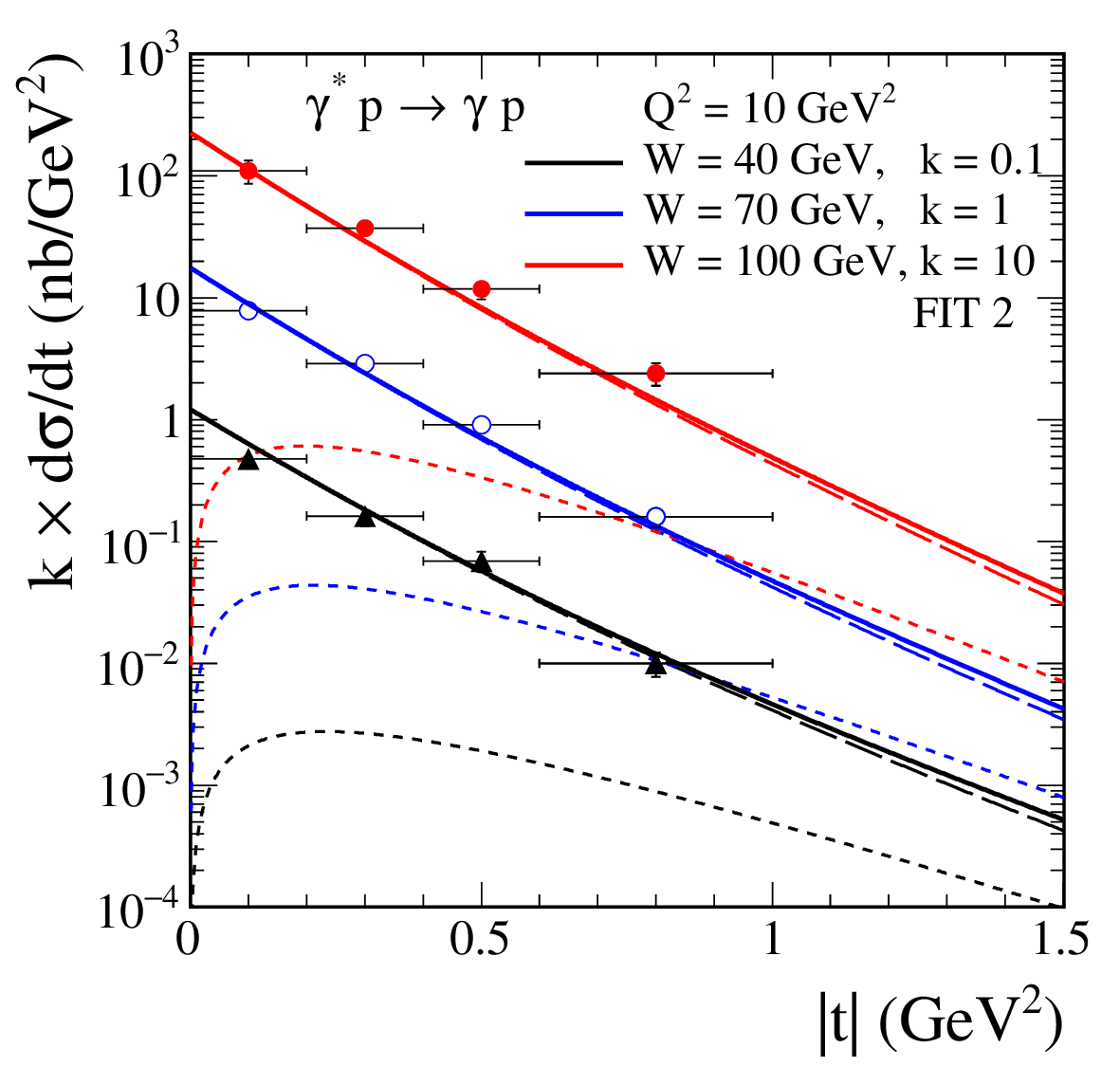}
\caption{\label{fig:3}
\small
Top panels: The differential cross sections $d\sigma/dt$
compared to experimental data
for different $\gamma^{*} p$ c.m. energy
and photon virtuality.
The upper line corresponds to $W = 12.7$~GeV 
and $Q^{2} = 0$ and averaged FNAL data \cite{Breakstone:1981wm}.
The theoretical results for $\gamma^{*} p \to \gamma p$ 
at higher $W$ and $Q^{2}$ 
for transverse (long-dashed lines) and longitudinal (short-dashed lines)
polarization of the $\gamma^{*}$ and their sum (solid lines)
are shown together with experimental data.
Data for $\langle W \rangle = 104$~GeV are from \cite{ZEUS:2008hcd}
and for $\langle W \rangle = 82$~GeV from \cite{H1:2009wnw}.
Bottom panels: 
Comparison of the fit results
to the H1 data \cite{H1:2009wnw}
for $\langle Q^{2} \rangle = 10$~GeV$^{2}$ 
and $\langle W \rangle = 40$, 70, and 100~GeV
(from bottom to top).
Here we show the results scaled by a factor $\rm k$
(specified in the figure legend)
for displaying purposes.}
\end{figure}

In Fig.~\ref{fig:3} we show
the differential cross sections $d\sigma/dt$
from (\ref{3.100}) with $\varepsilon = 1$
for different $\langle W \rangle$ and $\langle Q^{2} \rangle$.
In the top panels,
the upper line corresponds to $W = 12.7$~GeV 
and $Q^{2} = 0$ and should be compared 
to the averaged FNAL data (top data points)
for the $\gamma p \to \gamma p$ reaction
\cite{Breakstone:1981wm}.
In this kinematic range, for $Q^{2} = 0$
and at intermediate $W$, 
the reggeon plus soft-$\Pom$ contributions 
dominate and the hard-$\Pom$ exchange 
gives negligible contribution.
As expected, there is a significant interference between 
the reggeon and soft-pomeron components.
The slope parameters $b_{2}$ and $b_{1}$ in (\ref{t_dependence_ff})
are adjusted to the FNAL $d\sigma/dt$ data on 
the real-photon-proton scattering.
At higher $W$ and $Q^{2}$ measured at HERA
the hard pomeron plays an increasingly important role.
The slope parameter $b_{0}$ for the hard-pomeron exchange
is adjusted to the DVCS HERA data.
As we noted above, $d\sigma(\gamma^{*} p \to \gamma p)/dt$
is the sum of two contributions
$d\sigma_{\rm T}/dt$ and $d\sigma_{\rm L}/dt$ with the latter term
becoming very small for $|t| \to 0$.
This is understandable since for $\gamma^{*} p \to \gamma p$
forward scattering only double-helicity-flip amplitudes
can contribute to $d\sigma_{\rm L}/dt$.
Furthermore, $d\sigma_{\rm T}/dt$ is dominated 
by the $b$-type couplings
and $d\sigma_{\rm L}/dt$ is dominated by the $a$-type couplings;
see Eq.~(\ref{2.5}).

\begin{figure}[!ht]
\includegraphics[width=0.49\textwidth]{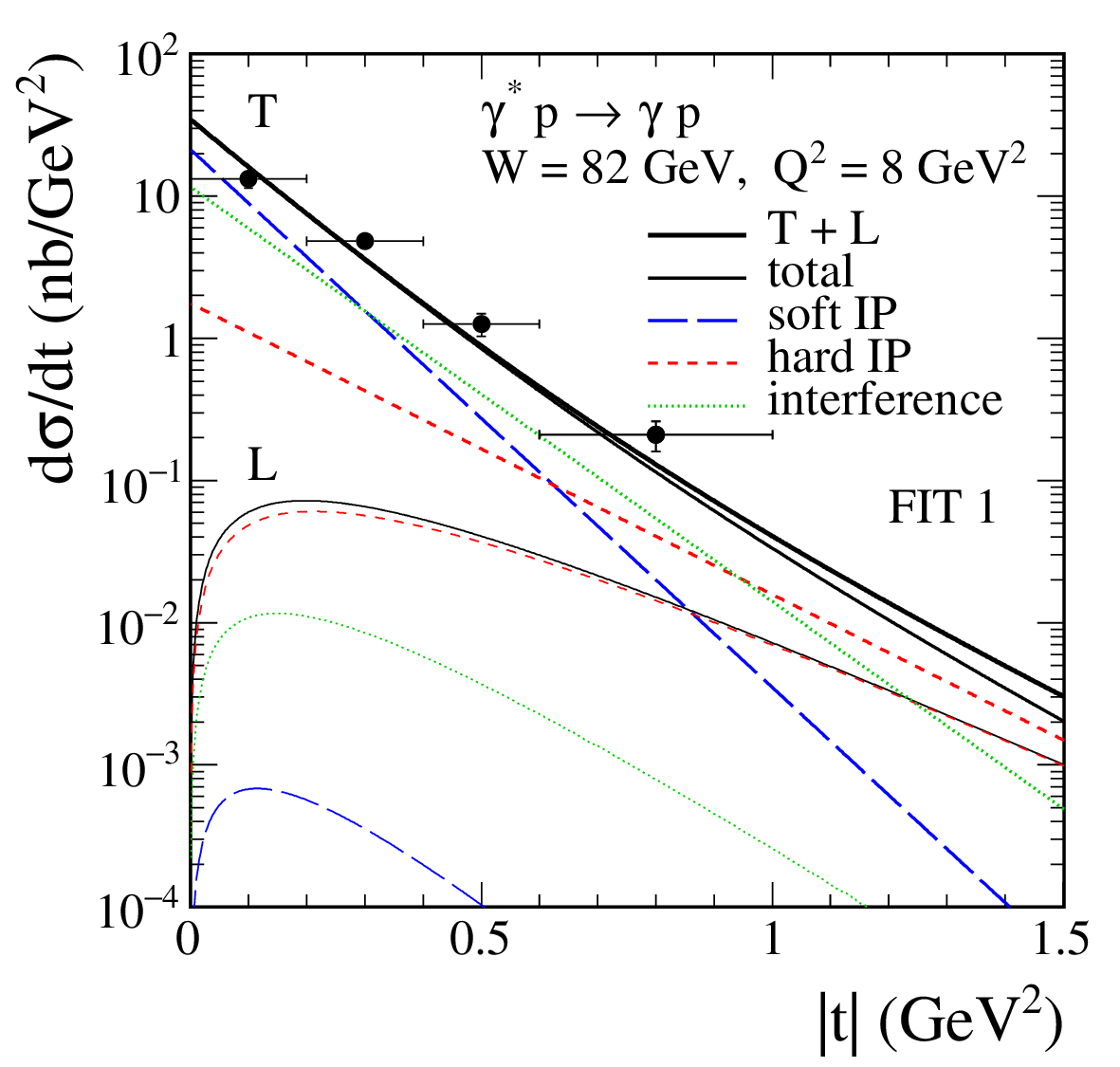}
\includegraphics[width=0.49\textwidth]{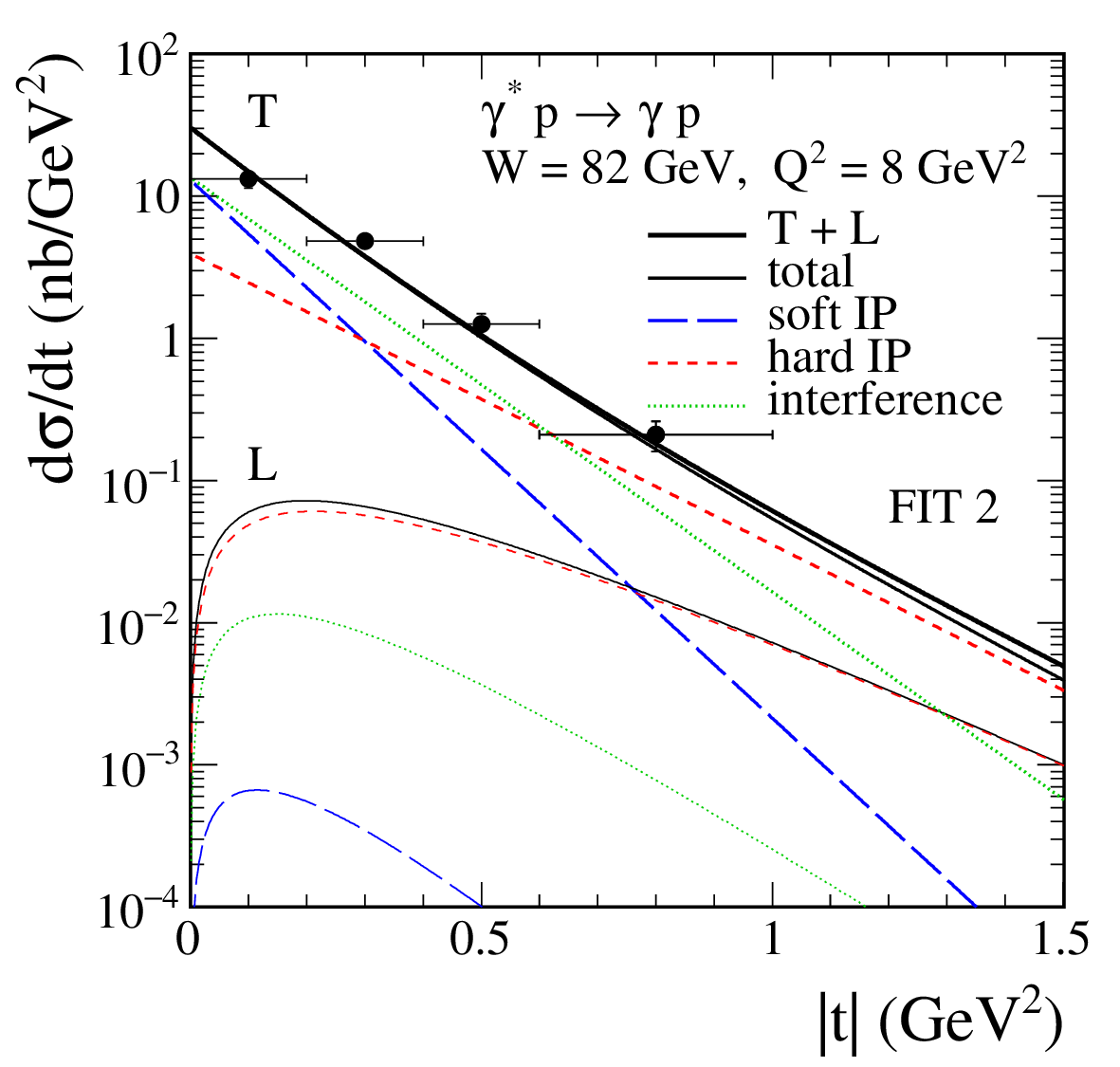}
\caption{\label{fig:4}
\small
Comparison of the FIT~1 (left) and the FIT~2 (right) 
to DVCS H1 data \cite{H1:2009wnw}
for $W = 82$~GeV and $Q^{2} = 8$~GeV$^{2}$.
The results for $\gamma^{*} p \to \gamma p$ 
for transverse (${\rm T}$) and longitudinal (${\rm L}$)
polarization of the $\gamma^{*}$ individually 
and their sum ${\rm T + L}$ (see the upper solid lines) are shown.
The contributions of soft $\Pom$ (the blue short-dashed lines), 
hard $\Pom$ (the red long-dashed lines),
the interference term (the green dotted lines),
and their sum total (the thin full lines)
for ${\rm T}$ and ${\rm L}$ components individually are also shown.}
\end{figure}

In Fig.~\ref{fig:4} we show the complete theoretical result
and individual components contributing to
the cross section $d\sigma/dt$, see (\ref{3.100}),
for $W = 82$~GeV and $Q^{2} = 8$~GeV$^{2}$
together with the H1 data \cite{H1:2009wnw}.
The constructive interference of the soft 
and hard pomeron terms is again a salient feature.

We note that our complete result for $d\sigma/dt$ 
does not have a simple exponential $t$ dependence 
as is frequently assumed in other models.
This is caused by the interference 
of soft- and hard-pomeron terms 
(each with different $t$ dependence)
and by the longitudinal contribution 
which is important for large $|t|$.

\begin{figure}[!ht]
\includegraphics[width=0.49\textwidth]{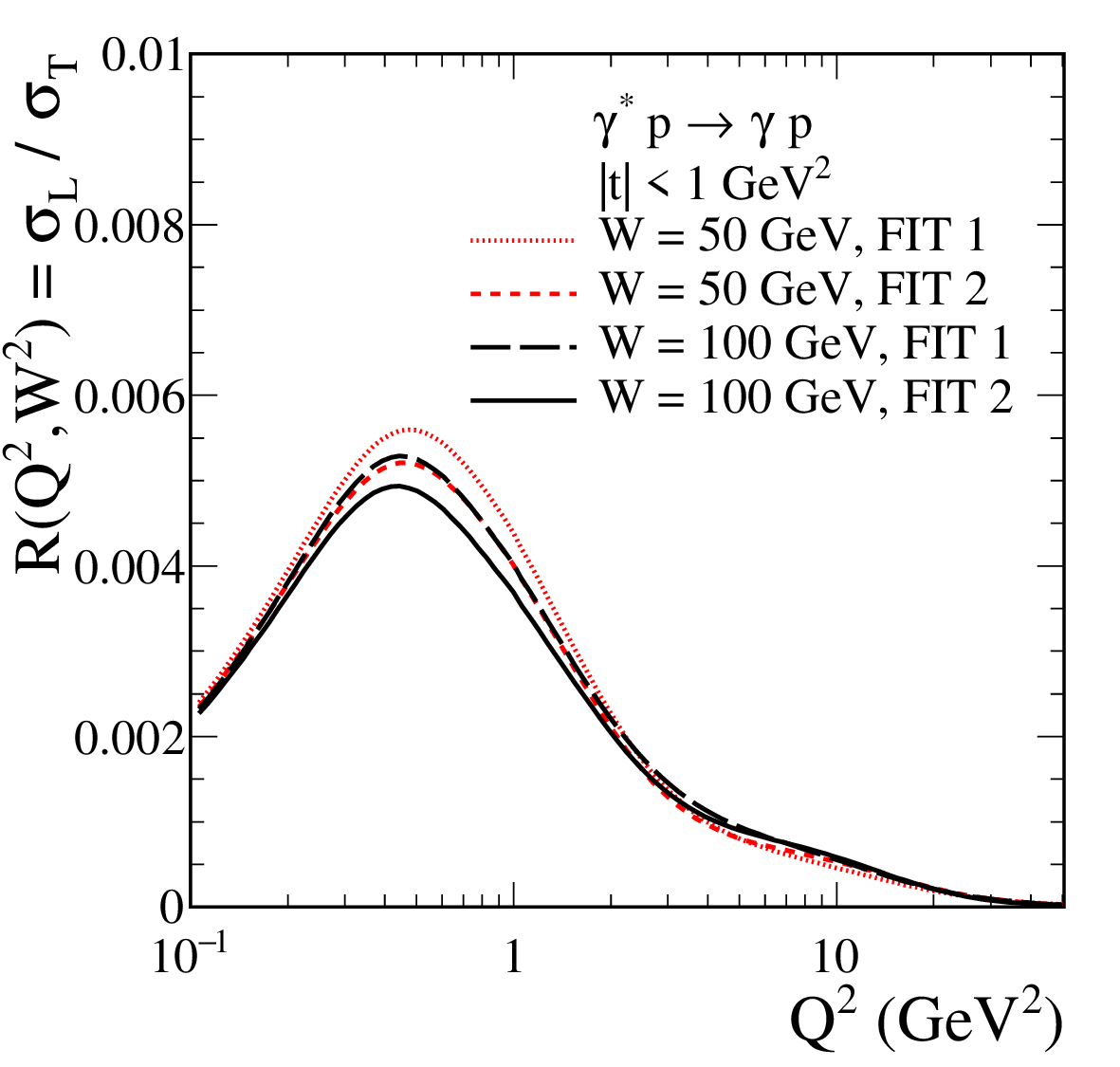}
\includegraphics[width=0.49\textwidth]{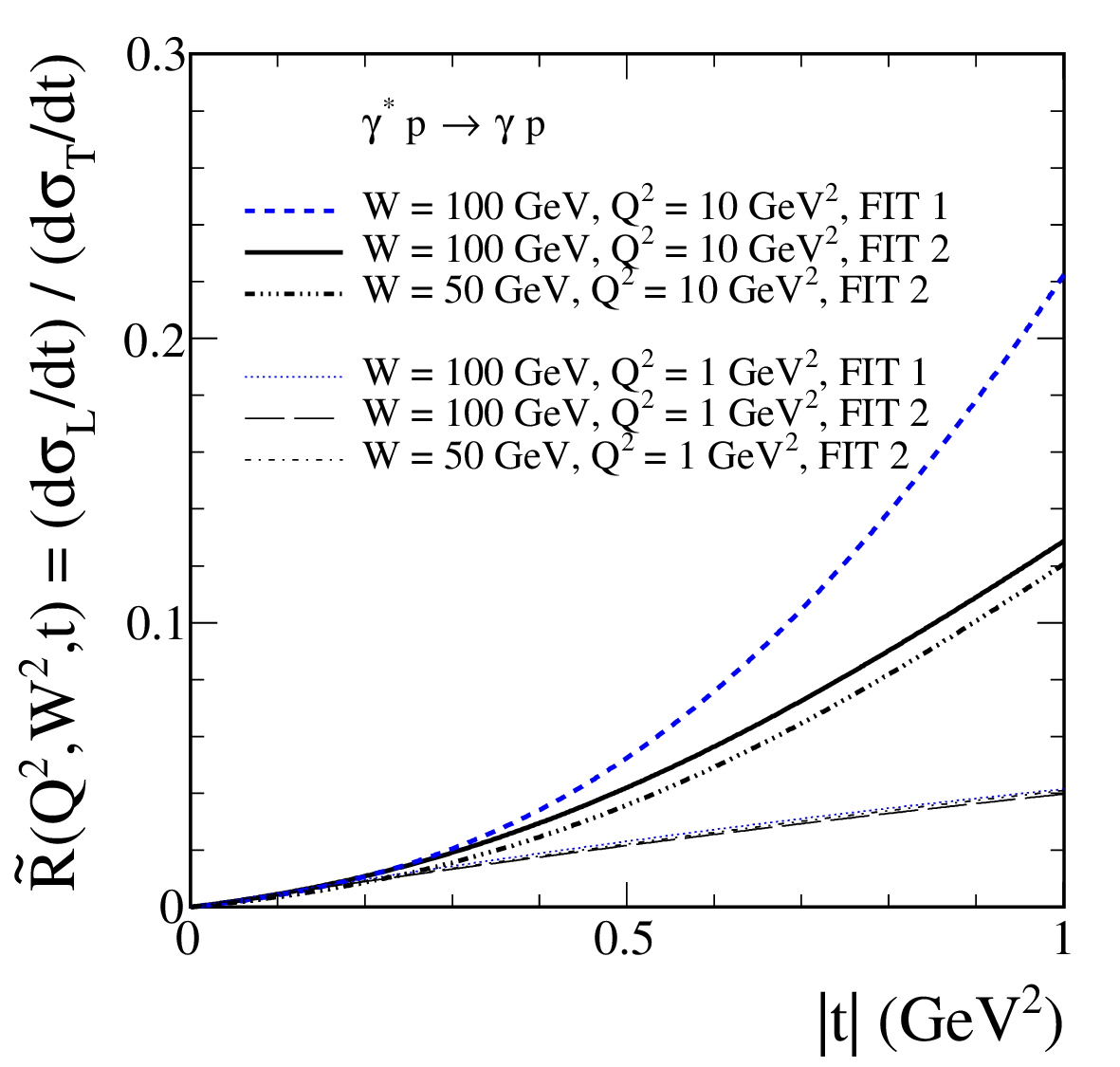}
\caption{\label{fig:ratios}
\small
The ratios of cross sections 
$R(Q^{2},W^{2})$ (\ref{ratio_Q2}) 
and $\tilde{R}(Q^{2},W^{2},t)$ (\ref{ratio_t})
for the $\gamma^{*} p \to \gamma p$ reaction
for longitudinally and transversely polarized 
$\gamma^{*}$.
Note that the meaning of the lines on these two panels is different.}
\end{figure}

Figure \ref{fig:ratios} shows the ratios
of the $\gamma^{*} p \to \gamma p$ cross sections
for longitudinally and transversely polarized
virtual photons,
\begin{eqnarray}
&&R(Q^{2},W^{2}) = \frac{\sigma_{\rm L}(Q^{2},W^{2})}{\sigma_{\rm T}(Q^{2},W^{2})} \,,
\label{ratio_Q2}\\
&&\tilde{R}(Q^{2},W^{2},t) = \frac{\frac{d\sigma_{\rm L}}{dt}(Q^{2},W^{2},t)}
                                  {\frac{d\sigma_{\rm T}}{dt}(Q^{2},W^{2},t)} \,,
\label{ratio_t}
\end{eqnarray}
as functions of $Q^{2}$ and $|t|$, respectively.
The cross section $\sigma_{\rm L}$
vanishes proportionally to $Q^{2}$ for $Q^{2} \to 0$.
The ratio $\tilde{R}(Q^{2},W^{2},t)$ strongly grows with $|t|$.
We must emphasize that this behaviour of 
$\tilde{R}(Q^{2},W^{2},t)$
depends crucially on our (reasonable) assumption
that $a$ and $b$ couplings in the $\Pom \gamma^{*} \gamma^{*}$
vertex functions have the same $t$ dependence 
for a given $j$~$(j = 0, 1, 2)$;
see (\ref{2.200}), (\ref{2.201}), and (\ref{t_dependence_ff}).
But it is clear from the right panel of Fig.~\ref{fig:ratios}
that a very small ratio $\tilde{R}(Q^{2},W^{2},t)$
for all $|t| \lesssim 1.0$~GeV$^{2}$ could, in our model,
only be  achieved if the $a$ and $b$ couplings
would have drastically different $t$ dependences.

\section{Conclusions and discussion}
\label{sec:conclusions}

In the present paper we have applied the tensor-pomeron
approach to deeply virtual Compton scattering 
(DVCS) for high c.m. energies $W$ 
and small Bjorken-$x$, $x \lesssim 0.02$.

We have made a comparison of the two-tensor-pomeron model 
to the DVCS data measured at HERA.
As a starting point we have used the fit parameters of the intercepts 
of the two pomerons and of the reggeon $\Reg_{+}$, 
and their coupling functions 
to real and virtual photons determined in \cite{Britzger:2019lvc}
from deep inelastic scattering (DIS) 
and real photoabsorption cross sections.
To our surprise, with a 'minimal', but reasonable, 
modification of the $Q^2$ dependence 
of only one $\gamma^*(Q^2) \gamma \Pom$ coupling function
[FIT~1 (\ref{FIT1})]
we have already got a good fit to the experimental data.
We could describe the $W$, $Q^2$ and $t$ dependences of
$d\sigma(\gamma^{*} p \to \gamma p)/dt$ measured at HERA
and of the elastic photon-proton cross section measured at FNAL.
The good description of the DVCS data 
is achieved due to a sizeable interference 
of soft and hard pomeron contributions.
We have considered also FIT~2 (\ref{FIT2}) 
in which the size of the hard-pomeron component was increased,
especially for larger $Q^2$,
and the soft-pomeron component was 
reduced relative to FIT~1.
We kept here, on purpose, the same parameters of the form factors (\ref{t_dependence_ff}) as in FIT~1.
The FIT~2 better describes the data 
at larger $|t|$ for $Q^2 \gtrsim 8$~GeV$^{2}$ 
(see Fig.~\ref{fig:3}).
Note that in our two-tensor-pomeron model 
the soft component and also the interference of soft and hard terms
are very important up to at least $Q^{2} \simeq 20$~GeV$^{2}$.

We consider it as a very satisfactory feature of our approach
for DVCS that we describe in the same framework
both the low $Q^{2}$ and high $Q^{2}$ regimes and the transition between them.
The same situation was shown to be true for DIS in \cite{Britzger:2019lvc}.
Going from reactions with virtual or real photons ($Q^{2} \leq 0$)
to hadronic reactions is straightforward in the tensor-pomeron model.
Indeed, in the original paper on the tensor-pomeron 
model~\cite{Ewerz:2013kda} hadronic reactions were the main focus, 
but already there photon-induced reactions were considered 
and general remarks on the vector-meson-dominance (VMD) model 
were made.
It can be seen from Refs.~\cite{Britzger:2019lvc,Ewerz:2013kda,
Lebiedowicz:2013ika,Bolz:2014mya,Lebiedowicz:2014bea,Ewerz:2016onn,
Lebiedowicz:2016ioh,Lebiedowicz:2019boz,Lebiedowicz:2019jru,
Lebiedowicz:2020yre,Lebiedowicz:2021byo,Lebiedowicz:2022nnn} 
and the present paper that
the tensor-pomeron model is applicable to soft high-energy
reactions which are purely hadronic as well as to reactions
involving photons and hadrons.

Now we comment on Refs.~\cite{Capua:2006ij,Fazio:2013hza} 
where also Regge theory is applied to DVCS. 
These authors consider, as far as we can see,
only leading helicity amplitudes.
For DVCS this means that only transversely polarized initial
$\gamma^{*}$'s are considered.
In contrast, in our tensor-pomeron approach 
we present a complete model,
including \textit{all} helicity amplitudes.
Therefore, we could make, e.g., predictions for
$\sigma_{\rm L}/\sigma_{\rm T}$ which can and should be checked
by experiments.
Also, our model for DVCS gives different relations between 
soft and hard terms compared to the analysis 
of \cite{Fazio:2013hza}, see Sec.~IV therein,
and also the reviews \cite{Jenkovszky:2018itd,Jenkovszky:2020jca}
where more results are presented.
The contribution from the interference term
found in \cite{Fazio:2013hza} is considerable
for intermediate values of $Q^{2}$, 
but smaller than our findings (FIT~1 and FIT~2).

In our calculations we have included the contributions 
of both transverse and longitudinal virtual photons. 
The longitudinal cross section $d\sigma_{\rm L}(Q^{2}, W^{2}, t)/dt$
is predicted to be very small for $|t| \to 0$ but
to be sizeable for 
$0.5 \; {\rm GeV}^{2} \lesssim |t| \lesssim 1.0\; {\rm GeV}^{2}$
(see Figs.~\ref{fig:3} and~\ref{fig:4}).
The corresponding ratio of ${\rm L/T}$ grows strongly with $|t|$
(see the right panel of Fig.~\ref{fig:ratios}).
We have also shown the $Q^2$ dependence of this ratio 
for different c.m. energies of the $\gamma^{*}p$ system.

To summarize, we have presented predictions for low-$x$ DVCS
of the two-tensor-pomeron model which previously
was successfully applied to low $x$ DIS in \cite{Britzger:2019lvc}.
The~model provides amplitudes for all helicity configurations
and, thus, can be checked by experimentalists in many ways.
We are looking forward to further tests of 
the non-perturbative QCD dynamics embodied in our
tensor-pomeron exchanges
in future electron-hadron collisions 
in the low-$x$ regime at the EIC and LHeC colliders.

\section*{Acknowledgments}
The authors are grateful to Markus Diehl for
correspondence and comments and to Carlo Ewerz 
for useful discussions.
This work was partially supported by
the Polish National Science Centre Grant
No. 2018/31/B/ST2/03537.
P. L. was supported by
the Bekker Program of the Polish National Agency for Academic Exchange, Project No. BPN/BEK/2021/2/00009/U/00001.


\end{document}